\documentclass[a4paper,fleqn,usenatbib]{mnras}
\usepackage[T1]{fontenc}
\usepackage{ae,aecompl}

\usepackage{graphicx}	
\usepackage{amsmath}	
\usepackage{amssymb}	

\usepackage{ulem}

\usepackage{xcolor}
\usepackage{ragged2e}

\title[GRMHD flow around black hole]
	{Study of general relativistic magnetohydrodynamic accretion flow around black holes }

\author[Mitra et al.]{Samik Mitra$^{1}$\thanks{E-mail: m.samik@iitg.ac.in (SM)}, 
Debaprasad Maity$^{1}$\thanks{E-mail: debu@iitg.ac.in (DM)}, Indu Kalpa Dihingia$^{2}$, Santabrata Das$^{1}$\thanks{E-mail: sbdas@iitg.ac.in (SD)}\\
$^1$ Indian Institute of Technology Guwahati, Guwahati, 781039, Assam, India\\
$^2$ Discipline of Astronomy, Astrophysics and Space Engineering, Indian Institute of Technology Indore, Indore 453552, India
}

\date{Accepted XXX. Received YYY; in original form ZZZ}

\begin{document}
\label{firstpage}
\pagerange{\pageref{firstpage}--\pageref{lastpage}}
\maketitle

\begin{abstract}
		We present a novel approach to study the global structure of steady, axisymmetric, advective, magnetohydrodynamic (MHD) accretion flow around black holes in full general relativity (GR). Considering ideal MHD conditions and relativistic equation of state (REoS), we solve the governing equations to obtain all possible smooth global accretion solutions. We examine the dynamical and thermodynamical properties of accreting matter in terms of the flow parameters, namely energy (${\cal E}$), angular momentum (${\cal L}$), and local magnetic fields. For a vertically integrated GRMHD flow, we observe that toroidal component ($b^\phi$) of the magnetic fields generally dominates over radial component ($b^r$) at the disk equatorial plane. This evidently suggests that toroidal magnetic field indeed plays important role in regulating the disk dynamics. We further notice that the disk remains mostly gas pressure ($p_{\rm gas}$) dominated ($\beta = p_{\rm gas}/p_{\rm mag} > 1$, $p_{\rm mag}$ refers magnetic pressure) except at the near horizon region, where magnetic fields become indispensable ($\beta \sim 1$). We observe that Maxwell stress is developed that eventually yields angular momentum transport inside the disk. Towards this, we calculate the viscosity parameter ($\alpha$) that appears to be radially varying. In addition, we examine the underlying scaling relation between $\alpha$ and $\beta$, which clearly distinguishes two domains coexisted along the radial extent of the disk. Finally, we discuss the utility of the present formalism in the realm of GRMHD simulation studies.
	
\end{abstract}

\begin{keywords}
accretion, accretion disc - black hole physics - (magnetohydrodynamics) MHD - magnetic fields
\end{keywords}


\section{Introduction} 
\label{sec:intro}

Black hole X-ray binary sources (BH-XRBs) are often considered to be the most ideal cosmic laboratory to probe the effect of strong gravity due to their rapid dynamical evolution in millisecond time scale \cite[and references therein]{Belloni-etal2000,Belloni-etal2005,Remillard-McClintock2006,Nandi-etal2012,Nandi-etal2018,Sreehari-etal2019,Baby-etal2020,Baby-etal2021,Majumder-etal2022}.  These BH-XRBs are embedded in the disk of inwardly spiralling accreting matter that are known to emit X-rays in the energy range of sub-keV to a few hundred keV \cite[]{Remillard-McClintock2006, Yuan-Narayan2014}. Therefore, the signatures of strong gravity are likely to be imprinted into those X-ray photons emitted from the surrounding of the BH-XRBs. Moreover, the existence of relativistic jets are observationally confirmed in both BH-XRBs and active galactic nuclei (AGN) \cite[]{Curtis1918,Jennison-DasGupta1953, Baade-Minkowski1954,Zensus1997,Davis-Tchekhovskoy2020,Janssen-etal2021}, which are presumed to be launched from the vicinity of the black hole \cite[and references therein]{Blandford-Znajek1977,Chakrabarti1999,Das-Chattopadhyay2008,Aktar-etal2015,Aktar-etal2017}. Indeed, the inflowing matter plays a viable role for the generation of such an enigmatic outflowing feature. In general, the differentially rotating convergent accretion flow around the central object is extremely viscous as well as turbulent \cite[]{Shakura-Sunyaev1973,Balbus-Hawley1991,Hawley-Balbus1995,Balbus-Hawley1998}. Since magnetic fields are ubiquitous in all astrophysical environments, the accretion flow around black hole is also expected to be indubitably magnetized in nature.

During the course of accretion, magnetic field is rooted in the disk either from the low-mass companion star or from the interstellar medium (\cite{Bisnovatyi-Kogan-Ruzmaikin1974,Bisnovatyi-Kogan-Ruzmaikin1976}), as these fields are `frozen in' to the accreting matter. In a magnetized disk, the dynamics of both inflowing and outflowing matters are primarily guided by the magnetic fields. In particular, the discovery of the magneto-rotational instability (MRI) suggests that accretion flows are driven by magnetohydrodynamical (MHD) turbulence \citep{Balbus-Hawley1991,Stone-etal1996,Balbus-Hawley1998,Hawley2000,Hawley-Krolik2001,Stone-Pringle2001,Pessah-etal2007} that eventually facilitates the angular momentum transport required to accrete matter onto the central object.

In studying the BH-XRBs and AGNs, the advection dominated accretion flow model received tremendous attention among the researchers \citep[for review]{Narayan-Yi1995,Yuan-Narayan2014}, although the magnetic fields are regarded there in stochastic limit. In reality, an accretion disk around black hole is likely to be threaded with large scale magnetic fields, and accordingly, several successive attempts were made to examine the accretion disk structure considering toroidal magnetic fields \citep{Akizuki-Fukue2006,Oda-etal2007,Begelman-Pringle2007,Oda-etal2010,Oda-etal2012,Samadi-etal2014,Sarkar-etal2018a,Sarkar-Das2018b, Dihingia-etal2020}. Meanwhile, \cite{Hirose-Krolik2004} reported that the ordered toroidal magnetic fields govern the flow dynamics at the inner part of the disk, whereas the plunging region is mostly dominant by the poloidal fields. Further, global MHD simulations (\cite{Hawley2001,Kato-etal2004}) revealed that inside the disk, the poloidal component of the magnetic fields remains weak compared to the toroidal one. \cite{Mishra-etal2020} examined the dynamical structure of geometrically thin accretion disks using global 3D MHD simulation and investigated the viscous effect resulted by means of the turbulent magnetic stress. \cite{Avara-etal2016} performed 3D general relativistic magnetohydrodynamic (GRMHD) simulations of radiatively efficient thin accretion discs and reported that large-scale magnetic field naturally accretes through the disk yielding enhanced radiative efficiency. Needless to mention that all these works are model dependent and hence, the plausible structure of the disk magnetic fields remains unsettled.

Very recently, Event Horizon Telescope collaboration (EHTC) performed a comprehensive analysis on the behaviour of the linear polarization of light emitted from M87$^*$ \cite[]{EHT-VII,EHT-VIII, Goddi-etal2021} and  for the first time, provided the novel insight of the magnetic field structures in the nearby region of any supermassive black hole (SMBH). The estimated magnetic field for M87$^*$ appears to be $1-30$ Gauss at $5r_g$, $r_g$ being the gravitational radius \cite[]{EHT-VII,EHT-VIII} that yields $5-150$ Gauss at the horizon as obtained assuming a $1/r$ dependence \cite[]{Ripperda-etal2022}, $r$ being the radial coordinate. And, the observed polarization map possibly resulted due to ordered radial and/or vertical magnetic fields present in the emission region. Recently, simulation studies of magnetically arrested disk \cite[MAD;][]{Igumenshchev-etal2003,Narayan2003} successfully reproduced the similar polarimetric signatures \cite[]{Palumbo-etal2020,Narayan-etal2021,Yuan-etal2022}, which eventually indicate that the magnetic fields are dynamically important in the near-horizon region. Also, it is worth mentioning that the structure of seed magnetic fields plays an important role in governing the accretion-ejection mechanism around black hole. In particular, the initial magnetic fields affect the dynamical structure of the relativistic jets/outflows \cite[e.g.][]{Nathanail-etal2020,Dihingia-etal2021}. This evidently indicates that the choice of the initial magnetic field structure is important but often it remains model dependent \cite[][]{Komissarov2006,Pu-etal2015} due the lack of steady-state GRMHD accretion solution available in the literature.

Being motivated with this, in this paper, we develop a formalism to study the MHD accretion flow in the general relativistic framework and provide an insight on the possible magnetic field configuration in the steady state. Here, we adopt ideal GRMHD approximations \cite[][]{Koide-etal1998, Koide-etal1999,Koide-etal2000, Koide2004,Mckinney-Gammie2004,McKinney2006} and ignore the exchange of energy between the plasmas and the radiation field (\cite{Anile1990, Porth-etal2019} and references therein) for simplicity. We consider vertically integrated MHD flow
and in this configuration, both radial ($b^r$) and toroidal ($b^\phi$) components of the magnetic fields are pertinent in regulating the accreting matter. With this, we solve the mass and energy-momentum conservation equations and obtain the complete set of global accretion solutions around Schwarzschild black hole. We calculate all relevant
dynamical and thermodynamical flow variables and study their dependence on flow parameters, such as energy (${\cal E}$), angular momentum (${\cal L}$) and local magnetic fields ($b^r$ and/or $b^\phi$). We observe that the magnetic field allows matter to accrete where toroidal component plays the dominant role in controlling the disk dynamics over the radial component. We notice that disk remain mostly gas pressure ($p_{\rm gas}$) dominated (plasma-$\beta=p_{\rm gas}/p_{\rm mag} > 1$, $p_{\rm mag}$ being magnetic pressure), however, magnetic fields become predominantly important at the near horizon region ($B \sim 10^6$ G at $r< 10 r_g$ for $M_{\rm BH}/M_\odot =1$, where $M_{\rm BH}$ is the mass of the black hole and $M_\odot$ is the solar mass). We further examine the viscous effect developed due to turbulent Maxwell stress and find that viscosity parameter ($\alpha$) is radially varying and often exceeds unity at the vicinity of the black hole. We also explore the nexus between $\alpha$ and plasma-$\beta$, and find two distinct scaling law features indicating the presence of separate accretion domains along the disk length. Overall, in this paper, for the first time to our knowledge, we provide a useful formalism to study the steady state global MHD accretion solutions around black hole in full general relativity.

The paper is organized as follows. In Section \ref{Space-time}, we present the model assumptions and governing GRMHD equations. In Section \ref{Solution}, we discuss the critical point analysis and the solution methodology. In Section \ref{sol-meth}, we elaborately present obtained results. Finally, we summarize our findings in Section \ref{Conclusion}.

\section{GRMHD formalism and underlying assumptions}

{\label{Space-time}}

We consider the general relativistic magnetohydrodynamic equations in a stationary axisymmetric spacetime. This spacetime possesses two commuting killing vectors associated with time ($t$) and azimuthal coordinate ($\phi$) and are given by $\xi^{t}$ and $\xi^{\phi}$, respectively. The general line element in this space-time is written in Boyer-Lindquist coordinate $(t,r,\theta, \phi)$  (\cite{Boyer-Lindquist1967}) as,
$$
\label{1}
ds^2 =
g_{tt}dt^2 + 2g_{t\phi}dtd\phi + g_{rr} dr^2 + g_{\theta\theta} d\theta^2 + g_{\phi\phi} d\phi^2.
\eqno(1)
$$ 
The BH is placed at the origin of the coordinate system and the event horizon is identified as $g^{rr}=1/g_{rr}=0$. We confine our calculations to the equatorial plane of the disk ($i.e.$, $\theta=\pi/2$). To express the flow variables, we use a unit system as, $M_{\rm BH} = G=c=1$, where $M_{\rm BH}$ is the black hole mass, $G$ is the gravitational constant and $c$ is the velocity of light, respectively. With this unit system, the radial coordinate, angular momentum, and flow velocity are measured in units of, $GM_{\rm BH}/c^2$, $GM_{\rm BH}/c$, and c, respectively. 

\subsection{GRMHD equations}

{\label{MHD}}

In order to describe the relativistic magnetized accretion processes, the governing GRMHD equations are obtained from the mass conservation, energy-momentum conservation and the homogeneous Faraday's law \cite[][]{Anile1990,DeVilliers-etal2003,Gammie-etal2003,Mckinney-Gammie2004} as follows,
	$$
	\nabla_\mu \left(\rho u^\mu \right) = 0;\qquad
	\nabla_\mu T^{\mu\nu}=0;\qquad
	\nabla_\mu {}^*F^{\mu\nu}=0.
	\eqno(2)
	$$ 
In these equations, $\rho$ is the mass density, $u^{\mu}$ is the four-velocity of matter, $T^{\mu\nu}$ is the stress energy-momentum tensor and $^{*}F^{\mu\nu}= \frac{1}{2}(-g)^{-1/2}\eta^{\mu\nu\delta\kappa}F_{\delta\kappa}$, denotes the Hodge dual of Faraday electromagnetic tensor $F^{\mu\nu}$. In general, the energy-momentum tensor is expressed as:
$$
T^{\mu\nu}_{Gen}=T^{\mu\nu}_{\rm FLU}+T^{\mu\nu}_{\rm VIS}+T^{\mu\nu}_{\rm MAX}+T^{\mu\nu}_{\rm RAD},
$$
where, $1^{\rm st}$, $2^{\rm nd}$, $3^{\rm rd}$ and $4^{\rm th}$ terms in the right hand side denote the contributions from the FLU-id, VIS-cous, MAX-well, and the RAD-iations \cite[][]{Abramowicz-Fragile2013}. For the purpose of simplicity, in this work, we restrict ourselves only with the FLU-id and MAX-well parts. With this, we obtain the simplified energy-momentum tensor as
	$$
	T^{\mu\nu}=(e+p_{\rm gas})u^\mu u^\nu + p_{\rm gas} g^{\mu\nu} + F^\mu_{~\lambda} F^{\nu \lambda} - \frac{1}{4} F^2 g ^{\mu\nu},
	\eqno(3)
	$$
where $e$ and $p_{\rm gas}$ are the internal energy density, and the gas pressure of the flow. Here, $F^2=F_{\mu \nu}F^{\mu \nu}$, and all the spacetime indices $(\mu,\nu,\lambda)$ run from $0 \rightarrow 3$.

In the fluid frame, $F_{\mu\nu}$ can be decomposed into electric field, $e^\mu = F^{\mu\nu}u_\nu$, and magnetic field $b^\mu = {}^*F^{\mu\nu}u_{\nu}$, such that the following relation holds \citep{Misner-etal1973,Baugmarte-Shapiro2003},
$$
F^{\mu\nu}=u^\mu e^\nu - u^\nu e^\mu - (-g)^{-1/2}\eta^{\mu\nu\lambda\delta}u_\lambda b_\delta.
\eqno(4)
$$
$$
^*F^{\mu\nu}=u^\mu b^\nu - u^\nu b^\mu - (-g)^{-1/2}\eta^{\mu\nu\lambda\delta}u_\lambda e_\delta.
\eqno(5)
$$
In this work, we consider ideal GRMHD approximation where conductivity of the fluid tends to infinity and consequently, electric field $e^\mu = 0$. This allows the magnetic field lines to remain frozen into the accreting plasmas that reduces the form of field tensors as,
$$
F^{\mu\nu}=-(-g)^{-1/2} \eta^{\mu\nu\lambda\delta}u_\lambda b_\delta,\\
{}^*F^{\mu\nu}= u^\mu b^\nu - u^\nu b^\mu.
\eqno(6)
$$

Using equation (6), we obtain the energy-momentum tensor as,
$$
T^{\mu\nu}=(e+p_{\rm gas})u^\mu u^\nu + p_{\rm gas} g^{\mu\nu} + \frac{1}{2}g^{\mu \nu} b^2+ b^2 u^\mu u^\nu - b^\mu b^\nu.
\eqno(7)
$$
After some simple algebra, we get,
$$
T^{\mu \nu}= \rho h_{\rm tot} u^\mu u^\nu + p_{\rm tot} g^{\mu \nu} - b^\mu b^{\nu}.
\eqno(8)
$$
Here, $h_{\rm tot}= h + B^2/\rho$, where the specific enthalpy of the fluid is given by $h=(e+p)/\rho$, and $p_{\rm tot} = p_{\rm gas} + p_{\rm mag}$ with $p_{\rm mag}=B^2/2$, where $1/\sqrt{4\pi}$ factor is absorbed while defining the magnetic fields.
The square of the magnetic field strength measured in the fluid frame is computed as $B^2= b_\mu b^\mu$.

\subsection{Conserved quantities in GRMHD}

{\label{Conserved}}

From the particle number conservation the continuity equation boils down to,
$$
\sqrt{-g} \rho u^r = {\rm constant.}
\eqno(9)
$$
The energy-momentum conservation supplemented by the killing condition, $\nabla_\mu \xi_\nu + \nabla_\nu \xi_\mu = 0$, assumes the following form,
$$
\nabla_\mu (T^{\mu \nu} \xi_{\nu})=0.
\eqno(10)
$$
The above equation (10) will provide us two conserved quantities which are as follows,
$$
\frac{-\sqrt{-g} \hspace{0.3cm}  T^r_t}{\sqrt{-g} \rho u^r}  = - h_{\rm tot} u_t +  \frac{1}{\rho u^r} b^r \big(g_{tt} b^t + g_{t\phi} b^\phi \big)=\mathcal{E},
\eqno(11)
$$
and
$$
\frac{\sqrt{-g} \hspace{0.3cm}  T^r_\phi}{\sqrt{-g} \rho u^r}= h_{\rm tot} u_\phi -  \frac{1}{\rho u^r} b^r \big(g_{\phi\phi} b^\phi + g_{t\phi} b^t \big)= \mathcal{L},
\eqno(12)
$$
where $\mathcal{E}$ and $\mathcal{L}$ are the globally conserved energy and angular momentum, respectively.

In addition, the time-component of source-free Maxwell's equation implies,
$$
\sqrt{-g} \mathcal{ B}^r=
\sqrt{-g} (u^t b^r - u^r b^t)= {\rm constant},
\eqno(13)
$$
and $\phi-$component equation implies the relativistic iso-rotation equation \citep{Mckinney-Gammie2004},
$$
\sqrt{-g} ~{}^*F^{r\phi}=\sqrt{-g} (u^r b^\phi - u^\phi b^r)= {\rm constant}.
\eqno(14)
$$
In the above equations the magnetic field components are expressed in terms of magnetic field 3-vector ($\mathcal{B}^i$) as,
$$
b^t = \mathcal{B}^i u^{\mu} g_{i\mu}, \qquad
b^i = \big(\mathcal{ B}^i + b^t u^i\big)/u^t.
\eqno(15)
$$
We construct the projection operator with respect to fluid frame as $\gamma^i_\mu = \delta^i_\mu + u^i u_\mu$, where $i$ runs from $1\rightarrow 3$. The projection operator also satisfies $\gamma^i_\mu u^\mu = 0$, which allows us to project the Navier-Stokes equation into three vector equations as
$$
\gamma^i_\mu \nabla_\nu T^{\mu\nu}=0.	
\eqno(16)
$$
Setting $i=r$ in equation (16), we obtain the radial momentum equation.

\subsection{Assumptions and Governing Equations}

{\label{Model}}

We consider a magnetized, advective accretion disc confined
around the black hole equatorial plane in the steady state. Therefore, given the background axisymmetry, we assume $\theta = \pi/2$ and consequently $u^\theta \sim 0$ throughout the disk. 
Further, we define the azimuthal velocity $v_\phi^2 = \frac{u^\phi u_\phi}{-u^t u_t}$ and the associated bulk azimuthal Lorentz factor as $\gamma_\phi^2=1/(1-v_\phi^2)$. Subsequently, the radial three-velocity in the corotating frame is defined
as, $v^2 = \gamma_\phi^2v_r^2$, where $v_r^2 = \frac{u^ru_r}{-u^tu_t}$ and the associated bulk Lorentz factor $\gamma_v^2 = 1/(1-v^2)$. Moreover, the specific angular momentum of the fluid is defined as, $\lambda = -u_\phi/u_t$ and the angular velocity is given by, 	
$$
\Omega = u^{\phi}/u^t = \frac{2 a_{\rm k} + \lambda (r-2)}{a_{\rm k}^2 (r+2) - 2 a_{\rm k} \lambda + r^3}.
\eqno(17)
$$

With this, the continuity equation (equation (9)) can be written in comoving frame as,
$$
\dot{M} = -4\pi \rho v \gamma_v  H \sqrt{\Delta},
\eqno(18)
$$
where ${\dot M}$ represents the accretion rate that we treat as global
constant. In the subsequent analysis, we express the accretion rate in terms of mass Eddington rate as $\dot{m} = \dot{M}/\dot{M}_{\rm Edd}$, where $\dot{M}_{\rm Edd}=1.44 \times 10^{18} \left(\frac{M_{\rm BH}}{M_{\odot}}\right)$ g $s^{-1}$. In this work, we choose $M_{\rm BH} = 1 M_\odot$ all throughout. In equation (18), $H$ denotes the local half-thickness of the disc which is calculated assuming the flow to be in hydrostatic equilibrium in the vertical direction and is given by \citep{Riffert-Herald1995,Peitz-Appl1997},
$$
H^2 = \frac{p_{\rm gas}r^3}{\rho \mathcal{F}}, \qquad \mathcal{F}=\gamma_\phi^2\frac{(r^2 + a_k^2)^2 + 2\Delta a_k^2}{(r^2 + a_k^2)^2 - 2\Delta a_k^2}.
\eqno(19)
$$

\subsubsection{Relativistic Equation of State}

{\label{REoS}}

The governing equations are closed with an equation of state (EoS) describing the relation among pressure ($p_{\rm gas}$), density ($\rho$), and internal energy ($e$). Following \cite{Chattopadhyay-Ryu2009}, we adopt an EoS for relativistic flow as
$$
e = \frac{\rho f}{\left(1+\frac{m_p}{m_e}\right)},
\eqno(20)
$$
with
$$
f=\left[ 1+ \Theta \left( \frac{9\Theta +3}{3 \Theta +2}\right) \right] + \left[ \frac{m_p}{m_e} + \Theta \left( \frac{9\Theta m_e +3m_p}{3 \Theta m_e+ 2 m_p}\right) \right],
$$
where $\Theta~(=k_{\rm B}T/m_e c^2)$ is the dimensionless temperature, $m_e$ is the mass of electron, and $m_p$ is the mass of ion. According to the relativistic EoS, we express the adiabatic index and polytropic index of the flow as $\Gamma = (1+N)/N$ and $N=(1/2)(df/d\Theta)$, respectively \citep{Dihingia-etal2019a}.

Following \cite{Gammie-etal2003}, the sound speed ($c_s$) and the Alfv\'en velocity ($c_a$) for relativistic flow are expressed as $c_s^2 = \Gamma p_{\rm gas}/\rho h$ and $c_a^2 = B^2/\rho h_{\rm tot}$, respectively. Moreover, \cite{Gammie-etal2003} introduces the dispersion relation for the fast MHD wave $\omega^2 = \left[c_s^2 + c_a^2 - c_s^2 c_a^2\right] k^2$, where $\omega$ and $k$ denote the frequency and the wavenumber of an MHD wave in the frame comoving with the fluid.  Accordingly, we obtain the Mach number $M =v/\sqrt{c_s^2 + c_a^2 - c_s^2 c_a^2}$ and the Alf\'venic Mach number $M_A =v/c_a$ of the flow.

\section{Critical point analysis/conditions}

Using equations (11), (12), (13), (14), (16) and (18), we obtain the wind equation of the flow (see Appendix \ref{Wind}) and is given by, 
$$
\frac{dv}{dr} = \frac{{\cal N}{(r,v,\Theta,\lambda, b^r, b^\phi)}}{{\cal D}{(r,v,\Theta,\lambda,b^r, b^\phi)}},
\eqno(21)
$$
where the numerator ${\cal N}$ and the denominator ${\cal D}$ are the explicit functions of $r$, $v$, $\Theta$, $\lambda$, $b^r,$ and $b^\phi$, and their expressions are given in Appendix \ref{Wind}. Similarly, the radial derivative of the other flow variables are expressed in terms of $\big(dv/dr\big)$ as,
$$
\frac{d\lambda}{dr} = \lambda_{11} + \lambda_{12} \frac{dv}{dr},
\eqno(22)
$$
$$
\frac{d\Theta}{dr} = \Theta_{11} + \Theta_{12} \frac{dv}{dr},
\eqno(23)
$$
$$
\frac{db^r}{dr} = b^r_{11} + b^r_{12} \frac{dv}{dr},
\eqno(24)
$$	
$$
\frac{d b^\phi}{dr} = b^\phi_{11} + b^\phi_{12} \frac{dv}{dr}.
\eqno(25)
$$
The explicit expressions of the coefficients, namely $\lambda_{11}$, $\lambda_{12}$, $\Theta_{11}$, $\Theta_{12}$, $b^r_{11}$, $b^r_{12}$, $b^\phi_{11}$, and, $b^\phi_{12}$ are given in Appendix. 

{\label{Solution}}

During the course of accretion around the black hole, the inflowing matter starts its journey from the disk outer edge ($r_{\rm edge}$) with negligible radial velocity (subsonic) and ultimately, enters into the black hole satisfying infall boundary conditions at the horizon ($r_h$). Because of this, accretion flow around black hole must change its sonic state at the critical point ($r_c$) and becomes transonic at least once, if not more. Such points are located in between $r_h$ and $r_{\rm edge}$. 
At the critical point, equation (21) has the form $(dv/dr)_{r_c}=0/0$ as both numerator (${\cal N}$) and denominator (${\cal D}$) simultaneously vanish there, and we have the critical point conditions ${\cal N}_{r_c}={\cal D}_{r_c}=0$. Accordingly, we apply the l$^{\prime}$Hospital rule to calculate $(dv/dr)_{r_c}$. In general, $(dv/dr)$ owns two distinct values at $r_c$: one is for accretion and the other is for wind. When both the values of $(dv/dr)_c$ are real and of opposite sign, the corresponding $r_c$ is called as saddle type critical point \citep[and references therein]{Matsumoto-etal1984, Kato-etal1993,Chakrabarti-Das2004}. Similarly, when $(dv/dr)_c$ are real, but of the same sign, $r_c$ is called as nodal type, and for imaginary values of $(dv/dr)_c$, the critical point becomes spiral type. In the astrophysical context, saddle type critical points have special importance as the global transonic accretion flow can only pass through it. Depending on the input parameters, GRMHD flow possesses either single or multiple critical points. When the critical point is formed near the horizon, it is referred as the inner critical point ($r_{\rm in}$), and when it forms far away from the horizon, we call them as outer critical point ($r_{\rm out}$) \cite[and references therein]{Chakrabarti-Das2004}. 

\section{Global accretion solutions}\label{sol-meth}

In order to obtain the global solution of the GRMHD accretion flow, one requires to solve the coupled differential equations (21-25) by employing the set of input parameters of the flow. Among these parameters, $\mathcal{E}$, $\mathcal{L}$, $a_{\rm k}$, and $\dot{m}$ are used as global parameters, whereas the critical point ($r_c$) and the radial magnetic field $b^r_c$ at $r_c$ are treated as local parameters. In this work, we consider flows around static black holes with Kerr parameter $a_{\rm k}=0$ and also set ${\dot m}=0.01$ all throughout unless stated otherwise. Using these flow parameters, we simultaneously solve ${\cal N} = 0$, and ${\cal D} = 0$ to calculate the radial velocity ($v_c$), temperature ($\Theta_c$), specific angular momentum ($\lambda_c$) and toroidal magnetic fields ($b^\phi_c$) at $r_c$. Employing these parameters, we first integrate equation (21) inwards up to the horizon and then outwards up to a large distance, equivalently the outer edge of the disk ($r_{\rm edge} \sim 1000$). Subsequently, we join both segments of the solutions to obtain the global transonic accretion solutions around the black holes. Depending on the input parameters, accretion flow passes through either the inner critical point ($r_{\rm in}$, usually forms close to the horizon), or outer critical point ($r_{\rm out}$, usually forms far away from the horizon) before entering into the black hole.

To this end, we emphasis that in the frame work of GRMHD, the accretion solutions passing through either inner critical point ($r_{\rm in}$) or outer critical point ($r_{\rm out}$) remain largely unexplored and hence, in this work, we intend to study the properties of the magnetized relativistic accretion flow around black holes extensively.

\subsection{Fluid properties of global accretion solutions containing inner critical point}

\begin{figure}
	\begin{center}
		\includegraphics[width=\columnwidth]{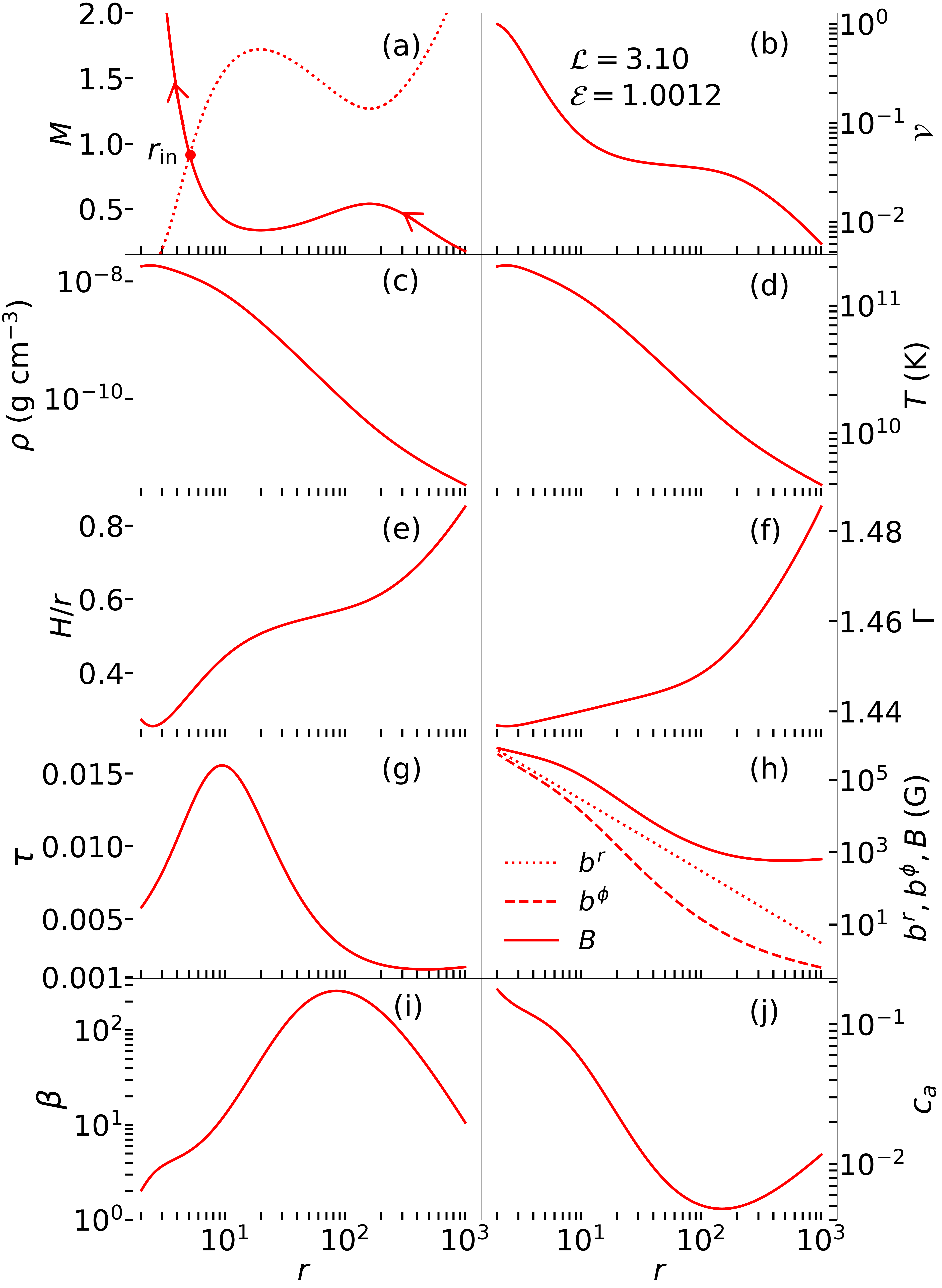}
		\caption{Example of a complete GRMHD accretion solution that passes through the inner critical point, $r_{\rm in}=5.1553$. Here, $\mathcal{E}=1.0012$, $\mathcal{L}=3.10$, and $b^r_{\rm in}=9.75 \times 10^4$ G are used. In panels (a)$-$(j), the profile of Mach number ($M$), velocity ($v$), density ($\rho$), temperature ($T$), disk aspect ratio ($H/r$), adiabatic index ($\Gamma$), vertical optical depth ($\tau$), magnetic field components ($B, b^r, b^\phi$), plasma-$\beta$, and Alf\'venic velocity ($c_a$) are plotted as function of radial distance ($r$). Filled circle denotes the location of $r_{\rm in}$ in panel (a). See text for details. 
		}
		\label{EL_B_sol}
	\end{center}
\end{figure}

In Fig. \ref{EL_B_sol}, we present a typical solution passing through the inner critical point ($r_{\rm in} = 5.1553$) where each panel shows the variation of the flow variables as function of the radial distance ($r$). This solution is obtained for $\mathcal{E}=1.0012$, $\mathcal{L}=3.10$, and $b^r_{\rm in}=9.75 \times 10^4$ G that smoothly connects the black hole horizon with the outer edge of the disc $r_{\rm edge} = 1000$. In Fig. \ref{EL_B_sol}a, we present the Mach number ($M$) variation of the transonic flow solutions. In this work, our interest is to focus only on the accretion solution (solid curve), however, for the purpose of completeness, we demonstrate its corresponding wind branch (dotted curve) as well. We observe that sub-sonic accretion flow from the outer edge of the disk ($r_{\rm edge} = 1000 r_g$) gradually gains its radial velocity as it moves inwards and eventually makes smooth transition to become super-sonic at the inner critical point ($r_{\rm in}=5.1553$) before falling into the black hole. At $r_{\rm in}$, we obtain the other flow variables as $v_{\rm in} = 0.1999$, $\Theta_{\rm in} = 27.9544$, $\lambda_{\rm in} = 3.1717$ and $b^\phi_{\rm in} = 6.73 \times 10^4$ G. In the figure, arrows indicate the direction of the flow motion and inner critical point ($r_{\rm in}$) is marked using filled circle. In Fig. \ref{EL_B_sol}b, we show the radial velocity ($v$) variation of the flow corresponding to the accretion solution depicted in Fig. \ref{EL_B_sol}a and find that flow enters into the black hole with velocity comparable to the speed of light. We demonstrate the density profile of the accreting flow in Fig. \ref{EL_B_sol}c, where gradual increase of density is observed as the flow proceeds towards the black hole. This happens mainly due to the geometric compression of the flow, and as a consequence, temperature of the flow is also increased with the decrease of radial distance as shown in Fig. \ref{EL_B_sol}d. We find that the disk becomes sufficiently hot with temperature as large as $T \ge 10^{11}$ K at the near horizon region with $r < 8 r_g$. We display the dependence of the vertical scale height ($H/r$) on the radial coordinate in Fig. \ref{EL_B_sol}e, where we find that $H/r$ remains less than unity all the way from the outer edge of the disc to the horizon. In Fig. \ref{EL_B_sol}f, we depict the profile of adiabatic index ($\Gamma$) as function of $r$. As expected, $\Gamma$ decreases with the decreasing $r$ and flow tends to become thermally trans-relativistic ($\Gamma \sim 1.4$) as it accretes towards the black hole \citep[and references therein]{Aktar-etal2015}. Further, we estimate the scattering optical depth $\tau = \kappa \rho h$, where the electron scattering opacity $\kappa = 0.38 ~{\rm cm}^2~{\rm g}^{-1}$ and present the obtained result in Fig. \ref{EL_B_sol}g. We observe that the flow remains optically thin ($\tau < 1$) even at the inner part of the disk ($r \lesssim 20 r_g$) although the density profile remains steeper there. This intuitively indicates that the possibility of escaping the high energy radiations from the inner part of the disk seems to be very much significant. In Fig. \ref{EL_B_sol}h, we display the variation of $b^r$ (dotted), $b^\phi$ (dashed) and $B$ (solid) with the radial distance. We find that although the strength of the magnetic fields is negligible ($b^r, b^\phi \sim 1$ G) at $r_{\rm edge}$, however, it is enhanced to $\sim 10^6$ G at the near horizon region yielding the inner part of the disk to be magnetically active. Next, we display the overall variation of the plasma-$\beta$ in panel Fig.\ref{EL_B_sol}i, where we find that disk remains mostly gas pressure dominated at all radii except at $r \lesssim 10 r_g$. Finally, we show the overall variation of the Alf\'venic velocity ($c_a$) in panel Fig.\ref{EL_B_sol}j, which initially decreases due to the slow increase of $B$, however enhances its value as flow moves towards the black hole. 

\subsection{General behaviour of global accretion solutions with fixed outer edge}

\begin{figure}
	\begin{center}
		\includegraphics[width=\columnwidth]{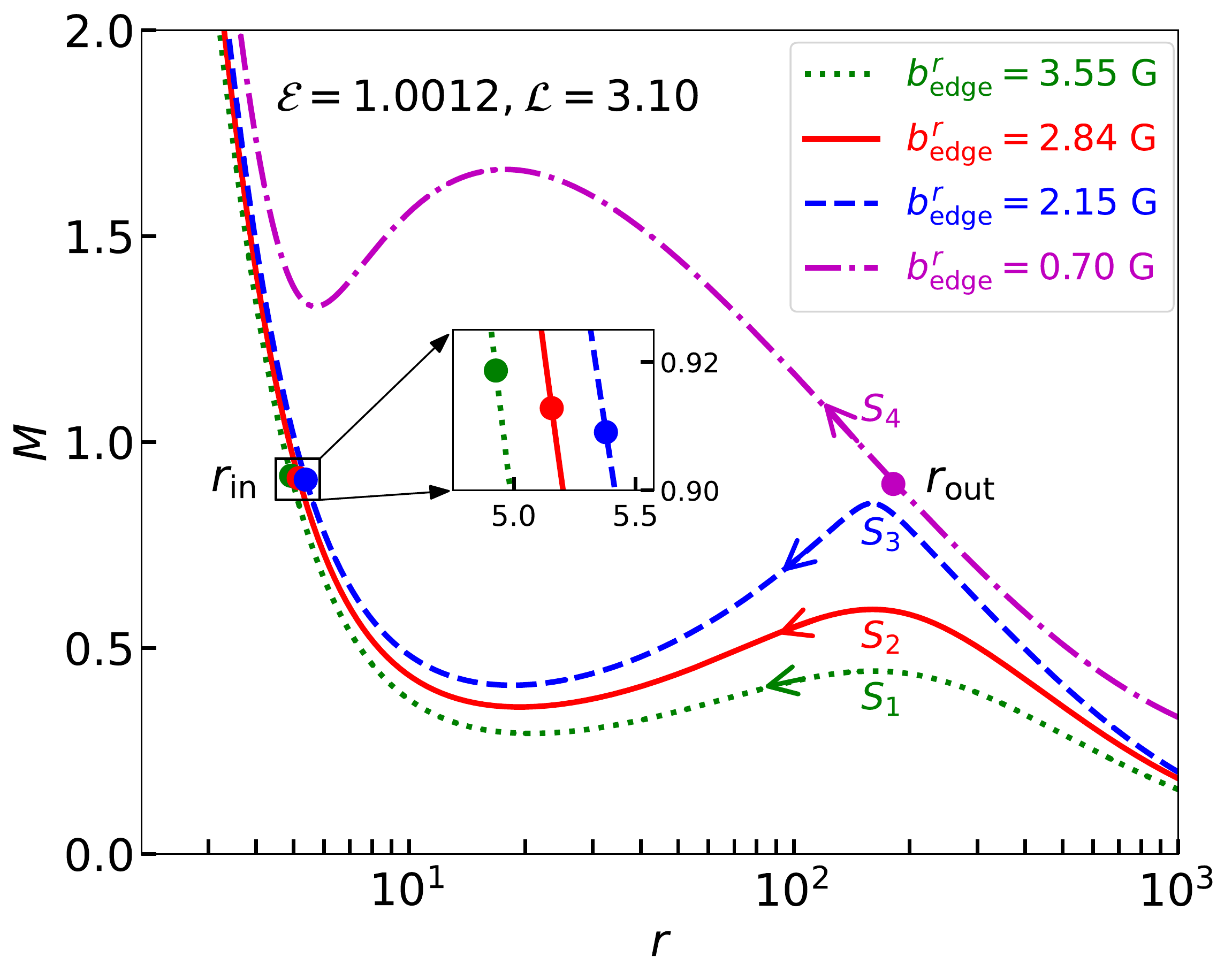}
		\caption{Variation of global GRMHD accretion solutions around black hole for different values of radial magnetic field ($b^r_{\rm edge}$) at the outer edge of the disk $r_{\rm edge}=1000$, where ${\cal E} = 1.0012$ and ${\cal L}=3.10$. The dotted ($S_1$, green), solid ($S_2$, red) and dashed ($S_3$, blue) curves denote the solutions for $b^r_{\rm edge}=3.55$ G, $2.84$ G, and $2.15$ G, respectively that pass through the inner critical points ($r_{\rm in}$). For the same set of the outer edge parameters, when $b^r_{\rm edge}=0.70$ G  is chosen, accretion solution passes through the outer critical point ($r_{\rm out}$) as depicted by dot-dashed ($S_4$, magenta) curve. In the figure, inner critical points are zoomed, and $r_{\rm in}$ and $r_{\rm out}$ are marked. Arrows indicate the direction of flow motion as it approaches towards the black hole. 
		}
		\label{M_r_in_out_Br}
	\end{center}
\end{figure}

In Fig. \ref{M_r_in_out_Br}, we examine the role of magnetic fields in deciding the nature of the accretion solutions having fixed outer boundary. Here, the dotted (green) curve demonstrates a global accretion solutions that starts its journey from $r_{\rm edge}=1000$ with $b^r_{\rm edge}=3.55$ G, ${\cal E} = 1.0012$, and ${\cal L}=3.10$, and it passes through the inner critical point $r_{\rm in}=4.9257$ with $b^r_{\rm in}=1.34 \times 10^5$ G before entering into the black hole. We mark this solution as $S_1$. Now, we decrease radial magnetic field to $b^r_{\rm edge}=2.84$ G keeping $\mathcal{E}$ and $\mathcal{L}$ unchanged, and calculate the global accretion solution by suitably tuning $v_{\rm edge}=0.006356$ and $\Theta_{\rm edge}=0.6680$. Here, we need to supply $v_{\rm edge}$ and $\Theta_{\rm edge}$ values additionally to obtain the accretion solution as the critical point is not known a priori. We plot this solution using solid (red) curve and for this solution $r_{\rm in}=5.1553$, and $b^r_{\rm in}=9.75 \times 10^4$ G. This solution is identical to the result presented in Fig. \ref{EL_B_sol} and marked as $S_2$. Upon decreasing $b^r_{\rm edge}$ gradually, we observe that below a minimum value of radial magnetic field at the outer edge $b^{r, {\rm min}}_{\rm edge}=2.15$ G, the accretion solution fails to pass through the inner critical point. For $b^{r, {\rm min}}_{\rm edge}=2.15$ G, the obtained accretion solution (marked as $S_3$) is shown by dashed (blue) curve, where $r_{\rm in}=5.3781$, and $b^r_{\rm in}= 6.78 \times 10^4$ G. When $b^r_{\rm edge} <b^{r, {\rm min}}_{\rm edge}$, namely $0.70$ G, the accretion solution changes its character allowing the flow to pass through the outer critical point ($r_{\rm out} = 181.465$) instead of inner critical point ($r_{\rm in}$) with $b^r_{\rm out}=21.08$ G, $\mathcal{E}=1.0012$ and $\mathcal{L}=3.10$. In the figure, this solution (marked as $S_4$) is depicted by the dot-dashed (magenta) curve. In the figure, filled circles denote the inner and outer critical points and arrows indicate the overall direction of the flow motion towards the black hole. 

\begin{figure}
	\includegraphics[width=\columnwidth]{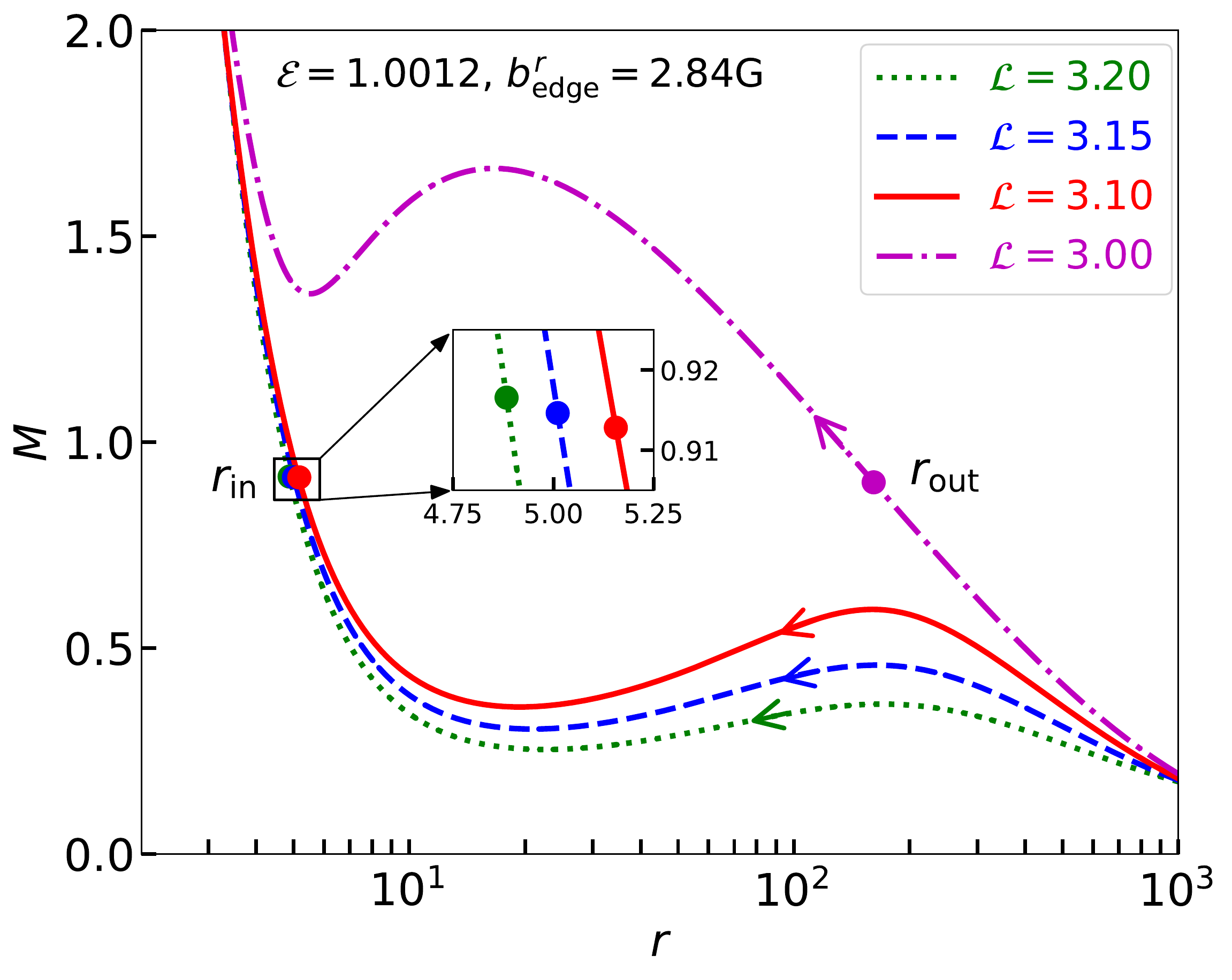}
	\caption{Similar to Fig. \ref{M_r_in_out_Br}, showing the variation of accretion solutions, when ${\cal E}=1.0012$ and $b^r_{\rm edge} = 2.84$ G are chosen and ${\cal L}$ is varied as marked in the figure. See text for details. 
	}
	\label{M_r_in_out_L}
\end{figure}

Since the nature of the transonic GRMHD accretion solutions also depend on ${\cal E}$ and ${\cal L}$, in addition to $b^r_{\rm edge}$, it is instructive to study their behavior by tuning these flow parameters. We find that the behavior of the accretion solutions changes as ${\cal L}$ is decreased for flows with ${\cal E}=1.0012$ and $b^r_{\rm edge} = 2.84$ G at $r_{\rm edge}=1000$. We present the obtained results in Fig. \ref{M_r_in_out_L}, where the solutions corresponding to ${\cal L} = 3.20$ (dotted, green), $3.15$ (dashed, blue), and $3.10$ (solid, red) are seen to pass through the inner critical points as $r_{\rm in}=4.8830$, $5.0105$, and $5.1553$, respectively,  whereas the solution with ${\cal L} = 3.0$ becomes transonic after crossing the outer critical point at $r_{\rm out} = 161.3239$ (dot-dashed, magenta). Thus, one arrives at conclusion that the effect of ${\cal L}$ is significant in deciding the nature of the accretion solutions around black holes. In the figure, the arrowed paths show the direction of the flow motion towards the black hole.

\begin{figure}
	\includegraphics[width=\columnwidth]{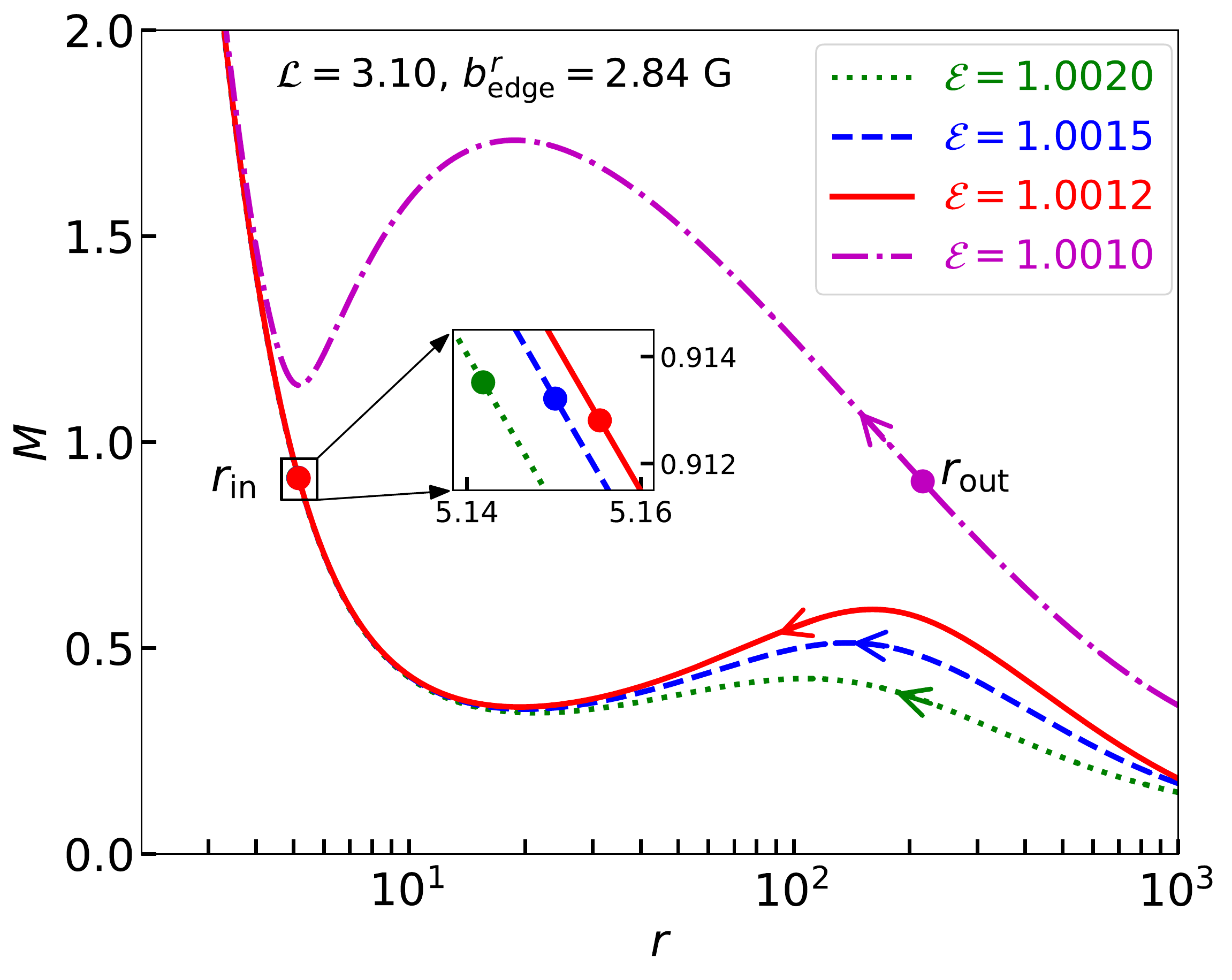}
	\caption{Similar to Fig. \ref{M_r_in_out_Br}, showing the variation of accretion solutions, when ${\cal L}=3.10$ and $b^r_{\rm edge} = 2.84$ G, and ${\cal E}$ is varied as marked in the figure. See text for details. 
	}
	\label{M_r_in_out_E}
\end{figure}

Similarly, Fig. \ref{M_r_in_out_E} gives the examples of accretion solutions that change their character due to the variation of ${\cal E}$. Here, we fix ${\cal L} = 3.10$, and $b^r_{\rm edge} = 2.84$ G at $r_{\rm edge}=1000$. The results plotted using dotted (green), dashed (blue), and solid (red) curves are for ${\cal E} = 1.0020$, $1.0015$ and $1.0012$, and these solutions cross the inner critical points at $r_{\rm in} = 5.1419$, $5.1502$, $5.1553$, respectively, before entering into the black hole. As the energy is decreased further, keeping all the remaining flow parameters unchanged, the accretion solution alters its trajectory and pass through the outer critical point at $r_{\rm out}=216.4050$ (dot-dashed, magenta) instead of the inner critical point. As before, we assert that ${\cal E}$ plays important role in stipulating the nature of the accretion solutions and arrows are used to indicate the direction of the flow motion.

\begin{figure}
	\includegraphics[width=\columnwidth]{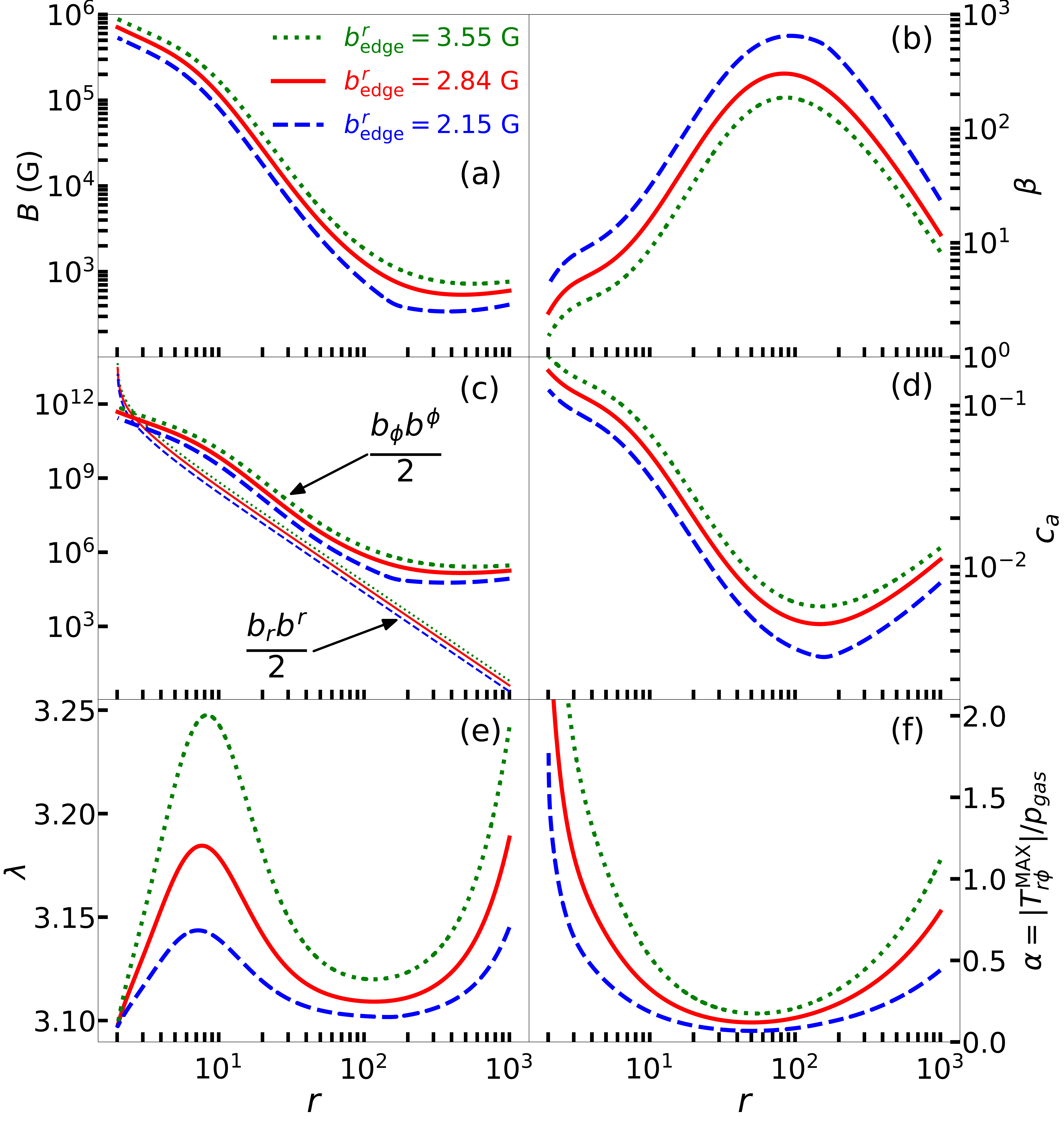}
	\caption{Parametric dependence of the flow variables for accreting matter around black hole. Various flow variables, namely (a) $B$, (b) plasma-$\beta$, (c) $b_r b^r/2$ and $b_\phi b^\phi/2$, (d) $c_a$, (e) $\lambda$, and (f) $\alpha$ are plotted as function of $r$. In each panel, dotted (green), solid (red), and dashed (blue) curves denote the results corresponding to the solutions marked as `$S_1$', `$S_2$' and `$S_3$' in Fig. \ref{M_r_in_out_Br}. See text for details. 
	}
	\label{BPBetaLambda_in}
\end{figure}

It is customary to examine the effect of magnetic fields on the properties of the disk fluid confined at the disk equatorial plane. For the purpose of representation, we consider accretion solutions marked as `$S_1$', `$S_2$' and `$S_3$' in Fig. \ref{M_r_in_out_Br} and present the associated fluid variables in Fig. \ref{BPBetaLambda_in}. While doing so, we show the variation of $B$ (Fig. \ref{BPBetaLambda_in}a), plasma-$\beta$ (Fig. \ref{BPBetaLambda_in}b), $b_r b^r/2$ and $b_\phi b^\phi/2$ (Fig. \ref{BPBetaLambda_in}c), Alf\'venic velocity (Fig. \ref{BPBetaLambda_in}d), specific angular momentum $\lambda$ (Fig. \ref{BPBetaLambda_in}e), and $\alpha$ (Fig. \ref{BPBetaLambda_in}f) with radial distance ($r$). In each panel, dotted (green), solid (red), and dashed (blue) curves are plotted for $b^r_{\rm edge} = 3.55$ G, $2.84$ G and $2.15$ G, respectively. As expected, we observe in panel (a) that the profile of $B$ corresponding to higher $b^r_{\rm edge}$ continues to remain higher compared to the cases with smaller $b^r_{\rm edge}$ values. It may be noted that in all cases, the flow starts with a very low radial magnetic fields ($b^r_{\rm edge} \sim 1$ G), however, the strength of the magnetic field tends to attain as high as $\sim 10^6$ G in the near horizon limit of the black hole. In panel (b), it is seen that as flow accretes towards the black hole, the gas pressure ($p_{\rm gas}$) initially increases compared to the magnetic pressure ($p_{\rm mag}$) leading to the increase of plasma-$\beta$. But, once the toroidal field component starts growing, plasma-$\beta$ decreases towards the black hole. Nevertheless, we find that disk is primarily gas pressure ($p_{\rm gas}$) dominated all throughout, although magnetic pressure ($p_{\rm mag}$) tends to become comparable to $p_{\rm gas}$ at the inner part of the disk. Needless to mention that $p_{\rm mag}$ is ascertained by the $b^r_{\rm edge}$ value; for a given $r$, higher $b^r_{\rm edge}$ renders enhanced $p_{\rm mag}$ as clearly seen in panel (b). Next, we illustrate the magnetic pressure corresponding to $r$ and $\phi$ components of the magnetic fields which are denoted by thin and thick curves in panel (c). We observe that the magnetic field strength is predominantly dominated by the toroidal component ($b^\phi b_\phi$) all throughout over the radial part ($b^r b_r$) except at the inner edge close to the horizon, $r \lesssim 3 r_g$. This finding is in agreement with the recent simulation work of \cite{Begelman-etal2022}. As observed before, higher $b^r_{\rm edge}$ yields enhanced magnetic pressure due to both $r$ and $\phi$ components. In panel (d), we see that relatively higher Alf\'venic velocity ($c_a$) is obtained for increasing $b^r_{\rm edge}$ values. Since $c_a$ is directly depends on the magnetic field strength, its radial variation in general follows the $B$ profile. In panel (e), we present how the specific angular momentum ($\lambda$) is transported in a magnetized accreting plasma. As the flow accrete towards the black hole from the outer edge of the disk, the effect of magnetic fields becomes increasingly important that causes the transport of angular momentum. In reality, the transport of $\lambda$ is mainly governed by the Maxwell stress ($T^{\rm MAX}_{r \phi} = B^2 u_r u_\phi - b_r b_\phi$), and therefore, it is evident that the profile of $\lambda$ strongly depends on the interplay among the flow variables. In general, for a given radial distance, $\lambda$ continues to remain higher for flows with larger $B$. However, the overall transport of $\lambda$ appears to be weak resulting the flow to remain sub-Keplerian all throughout the disk. This clearly indicates that MHD flow of this kind seems to remain weakly viscous all throughout the disk domain. Finally, in panel (f), we examine the profile of the viscosity parameter, $\alpha = |T^{\rm MAX}_{r \phi}|/p_{\rm gas}$, which is defined as the ratio of the Maxwell stress to the gas pressure \cite[and references therein]{Hawley-Krolik2001,Pessah-etal2007,Penna-etal2012,Mishra-etal2020}. We find that $\alpha$ varies with radial distance unlike the standard viscosity prescription of \cite{Shakura-Sunyaev1973}. As the flow proceeds inwards, $p_{\rm gas}$ initially increases over $p_{\rm mag}$ (see panel (b)) leading to the decrease of $\alpha$, although $\alpha$ eventually enhances its value once the magnetic stress starts to dominate. We notice that $\alpha$ exceeds its outer edge value at the inner part of the disk where magnetic fields are very high (see panel (a)) and because of this, rapid loss of $\lambda$ is observed at the vicinity of the black hole horizon. Overall, it is evident that $\alpha$ in magnetized disks varies with radial distance as it is computed using magnetic stress and this finding is in agreement with the results from GRMHD simulation \cite[and references therein]{Hawley-Krolik2002,Avara-etal2016}.

\begin{figure}
	\includegraphics[width=\columnwidth]{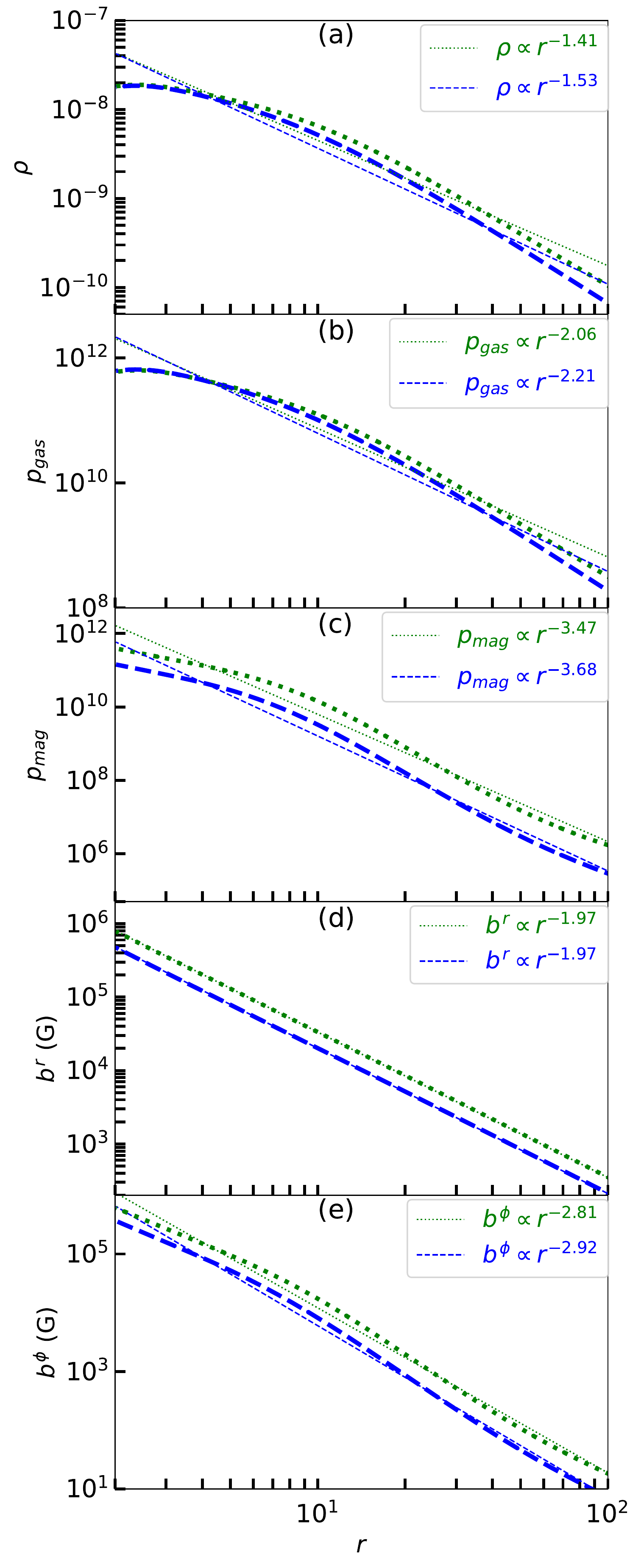}
	\caption{Best fitted power-law of (a) density $\rho$, (b) gas pressure $p_{\rm gas}$, (c) magnetic pressure $p_{\rm mag}$, (d) radial component of the magnetic fields $b^r$ and (e) toroidal component of the magnetic fields $b^\phi$. In each panel, thick dotted (green) and thick dashed (blue) curves are for solutions marked `$S_1$' and `$S_3$' in Fig. \ref{M_r_in_out_Br}, and thin dotted (green) and thin dashed (blue) correspond to the best fit power-law representations. See text for details. 
	}
	\label{power-law-fit}
\end{figure}

Now, we attempt to interpret the radial profiles of the following quantities, namely density ($\rho$), gas pressure ($p_{\rm gas}$), magnetic pressure ($p_{\rm mag}$), radial magnetic field ($b^r$), and toroidal magnetic field ($b^\phi$) by means of the power-law fit. For that we consider accretion solutions marked as `$S_1$' and `$S_3$' in Fig. \ref{M_r_in_out_Br}, and present the radial variation of the above quantities in Fig. \ref{power-law-fit}(a-e). In each panel, thick dotted (green) and thick dashed (blue) curves denote the results obtained from solutions `$S_1$' and `$S_3$', and the corresponding power-law fits are shown by thin dotted and thin dashed lines, respectively. We find that the best fit for density at all radii gives $\rho \propto r^{-(n+1/2)}$ (panel a) with $n \sim 1$, which seems to be consistent with the results of \cite{Narayan-Yi1995,Blandford-Begelman2004} for pure accretion having no outflow. Similarly, the best fits for the remaining quantities are obtained as $p_{\rm gas} \propto r^{-(n+7/6)}$, $p_{\rm mag} \propto r^{-(n+5/2)}$, $b^r \propto r^{-(n+1)}$, and $b^\phi \propto r^{-(n+9/5)}$. Needless to mention that  we observe in general poor fits of the accretion solutions in the near horizon limit. This possibly happens due to the fact that the transonic nature of the flow is not taken into account in the power-law fitting. 

\begin{figure}
	\includegraphics[width=\columnwidth]{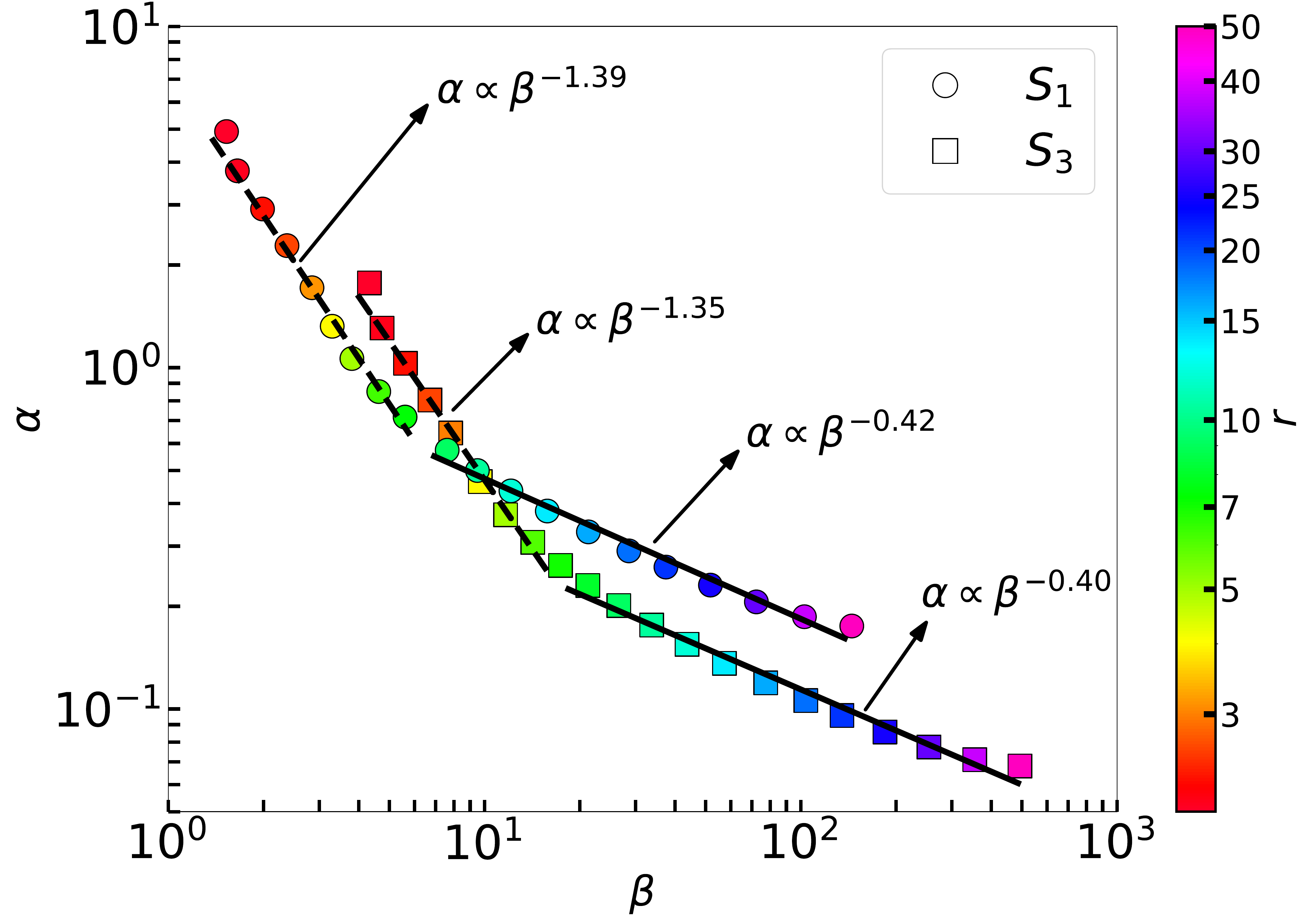}
	\caption{Viscosity parameter ($\alpha$) is plotted as a function of plasma-$\beta$ at all radii between $2 \le r \le 50$. Filled circles and filled squares are obtained for the accretion solutions marked `$S_1$' and `$S_3$' in Fig. \ref{M_r_in_out_Br}, whereas the solid and dashed lines (black) represent the best fitted power laws, respectively.	See text for details. 
	}
	\label{alp_beta}
	
\end{figure}

Next, we make an effort to reconcile our theoretical predictions with previous studies of magnetized accretion flow accomplished by the local shearing box simulations, where tight correlation between the plasma-$\beta$ and the viscosity parameter $\alpha$ is revealed as $\alpha \beta \sim 0.5$ \citep[and references therein]{Hawley-Balbus1995,Blackman-etal2008,Sorathia-etal2012,Salvesen-etal2016}. Towards this, in Fig. \ref{alp_beta}, we depict the correlation between $\alpha$ and $\beta$ for the accretion solutions marked `$S_1$' and `$S_3$' in Fig. \ref{M_r_in_out_Br}. Here, the results are obtained for $r \le 50 r_g$ just to collate with the existing simulation studies. In the figure, `$S_1$' and `$S_3$' solutions are plotted using filled circles and filled squares, respectively. The color code denotes the radial coordinate and its range is shown using colorbar at the right side of the figure. The best fit generally yields $\alpha \propto \beta^{-q}$, where two distinct domains are ascertained as a result of different exponents ($q$) values. For $6 r_g \lesssim r < 50 r_g$, we get the best fit value as $q \sim 0.4$ and $\sim 0.42$ corresponding to `$S_1$' and `$S_3$', respectively. This can be expressed approximately as $\alpha \propto \beta^{-2/5}$ which is in close agreement with the value $\sim 0.53$ as reported in \cite{Salvesen-etal2016}. In addition, at the inner part of the disk ($2 r_g < r \lesssim 6 r_g$) where the magnetic activity is relatively stronger, we obtain $q \sim 1.35$ and $\sim 1.39$ as depicted by dashed lines. Such stiff scaling relation (approximately $\alpha \propto \beta^{-7/5}$), to the best of our knowledge, has not yet been reported in the literature which we plan to take up for future works.

\subsection{Modification of accretion solutions possessing inner critical point}

\begin{figure}
	\includegraphics[width=\columnwidth]{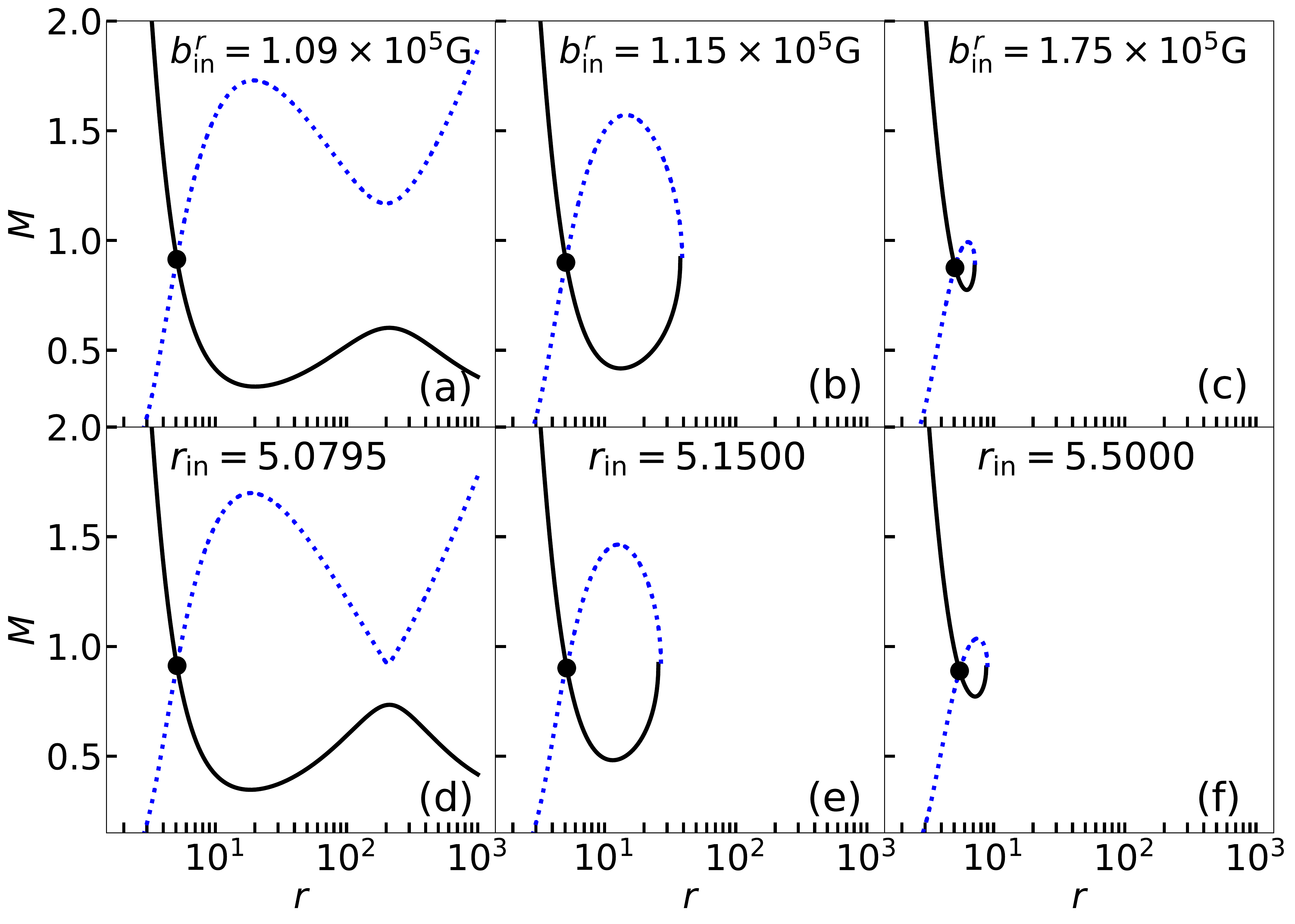}
	\caption{Plot of Mach number as a function of radial distance ($r$). Solid (black) curve denotes accretion solution and dotted (blue) curve refers corresponding wind solution. In the upper panels, we choose $r_{\rm in} = 5.0744$, ${\cal E} = 1.0012$, and ${\cal L}=3.10$, and increase the local radial magnetic field ($b^r_{\rm in}$) at $r_{\rm in}$ which are marked. In the lower panels, we consider $b^r_{\rm in} = 1.09 \times 10^5$ G, ${\cal E} = 1.0012$, and ${\cal L}=3.10$ and vary $r_{\rm in}$ as it is marked. See text for details. 
	}
	\label{M_r_rin_br_vary}
\end{figure}

In this section, we examine how the nature of the accretion solutions alters due to the variation of either magnetic fields or inner critical points for flows with a given set of (${\cal E}$, ${\cal L}$) values, and plot them in Fig. \ref{M_r_rin_br_vary}. In each panels of the figure, the Mach number ($M$) is depicted as function of radial distance ($r$), where solid (black) and dotted (blue) curves represent the accretion and wind solutions, respectively and filled circles denote the inner critical points. In the upper panels, we choose $r_{\rm in} = 5.0744$, ${\cal E} = 1.0012$, and ${\cal L}=3.10$, and vary radial component of the magnetic fields as (a) $b^r=1.09 \times 10^5$ G, (b) $1.15 \times 10^5$ G, and (c) $1.75 \times 10^5$ G. In panel (a), the flow passes through the inner critical point and smoothly connects the event horizon to the outer edge of the accretion disc ($r_{\rm edge}=1000$). When the radial magnetic field is increased to $b^r=1.15 \times 10^5$ G  (panel (b)), the flow solution becomes closed in the range $r_{\rm in} < r < r_{\rm out}$ with Mach number $M(r) = M_c$ \citep{Chakrabarti-Das2004} and fails to connect the black hole horizon with the outer edge of the disc ($r_{\rm edge}$). However, this solution can join with another solution passing through the outer critical point, if it exists, via shock transition \citep{Fukue1987,Chakrabarti1989,Das2007}, and accordingly, the accretion solution can extends up to $r_{\rm edge}$. In reality, this happens because the inner critical point solution possesses higher entropy than the outer critical point solution \citep{Becker-Kazanas2000}. It is noteworthy that the accretion solutions involving shocks are potentially promising in explaining the observational findings of the Galactic black hole sources \citep{Chakrabarti-Titurchuk1995,Chakrabarti-Manickam2000,Aktar-etal2015,Sreehari-etal2020,Das-etal2021}. The study of shock solutions for GRMHD flows is beyond the scope of the present paper and hence, will be reported elsewhere. As the radial magnetic field is increased further, the closed solution gradually shrinks and ultimately disappears as the critical point turns in to nodal type. With this, we indicate that for a given set of flow parameters, there exists two critical values of $b^r$ --- first one is the lower critical value for which the open solution passing through the fixed inner critical point becomes closed, and other one corresponds to the higher critical value for which saddle type critical points disappear. In the lower panels of Fig. \ref{M_r_rin_br_vary}, we choose flow parameters as $b^r_{\rm in}=1.09 \times 10^5$ G, ${\cal E} = 1.0012$, and ${\cal L}=3.10$, and vary the inner critical point as in panels (d) $r_{\rm in}=5.0795$, (e) $5.1500$, and (f) $5.5000$. 
Similar to the upper panels, we again find that as $r_{\rm in}$ is gradually receded away from the horizon, the flow behaviour changes their character from open type to closed type and ultimately it ceases to exist when $r_{\rm in}$ turns in to nodal type.

\begin{figure}
	\includegraphics[width=\columnwidth]{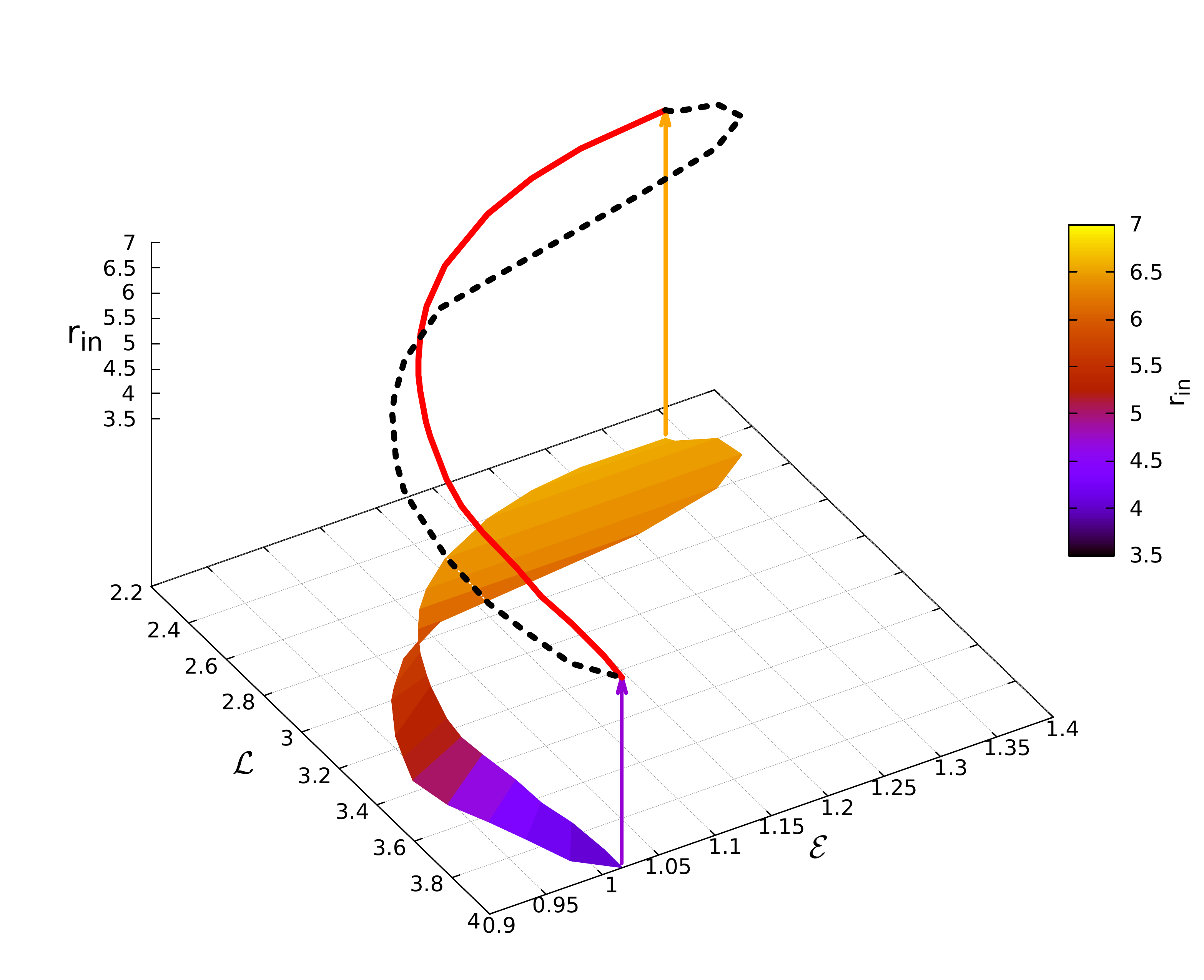}
	\caption{Plot of three-dimensional parameter space of flow energy ($\mathcal{E}$), angular momentum ($\mathcal{L}$), and inner critical point ($r_{\rm in}$). Here, we fix $b^r_{\rm in} = 1.15 \times 10^5$ G, $b^\phi_{\rm in}=1.00 \times 10^5$ G. Solid (red) and dashed (black) curves denote the boundary of the parameters surface. Two-dimensional surface projection of the three-dimensional plot is shown in ${\cal E}-{\cal L}$ plane, where color code denotes the range of $r_{\rm in}$. See text for details. 
	}
	\label{parasapce}
\end{figure}

So far, we have studied the global accretion solutions that pass through either inner or outer critical points. Here, we wish to emphasize that solutions of these kinds are not isolated solutions, instead they exist for a wide range of flow parameters. While envisaging this fact, we intend to examine the range of flow parameters that admit closed accretion solutions passing through the inner critical points (see Fig. \ref{M_r_rin_br_vary}). Accordingly, in Fig. \ref{parasapce}, we separate the effective domain of the parameter space spanned by $\mathcal{E}$, $\mathcal{L}$, and $r_{\rm in}$ that provides closed accretion solutions around black holes. Here, we fix $b^r_{\rm in}=1.15\times10^5$ G and $b^\phi_{\rm in}=1.00 \times 10^5$ G at $r_{\rm in}$ and plot the parameter space where solid and dotted curves denote its two edges. A wrapping of the the parameter space is clearly visible, which is possibly resulted due to the complex non-linearity involved among the GRMHD flow variables. Moreover, for the purpose of clarity, we present the two-dimensional projection of the three-dimensional parameter space in ($\mathcal{E},\mathcal{L}$) plane where color code denotes the allowed range of $r_{\rm in}$ as shown using colorbar. From the figure, it is evident that for smaller ${\cal L}$, generally higher $r_{\rm in}$ is required to obtain the closed accretion solution for GRMHD flow, and vice versa.

\subsection{Fluid properties of global accretion solution possessing outer critical point}

\begin{figure}
	\includegraphics[width=\columnwidth]{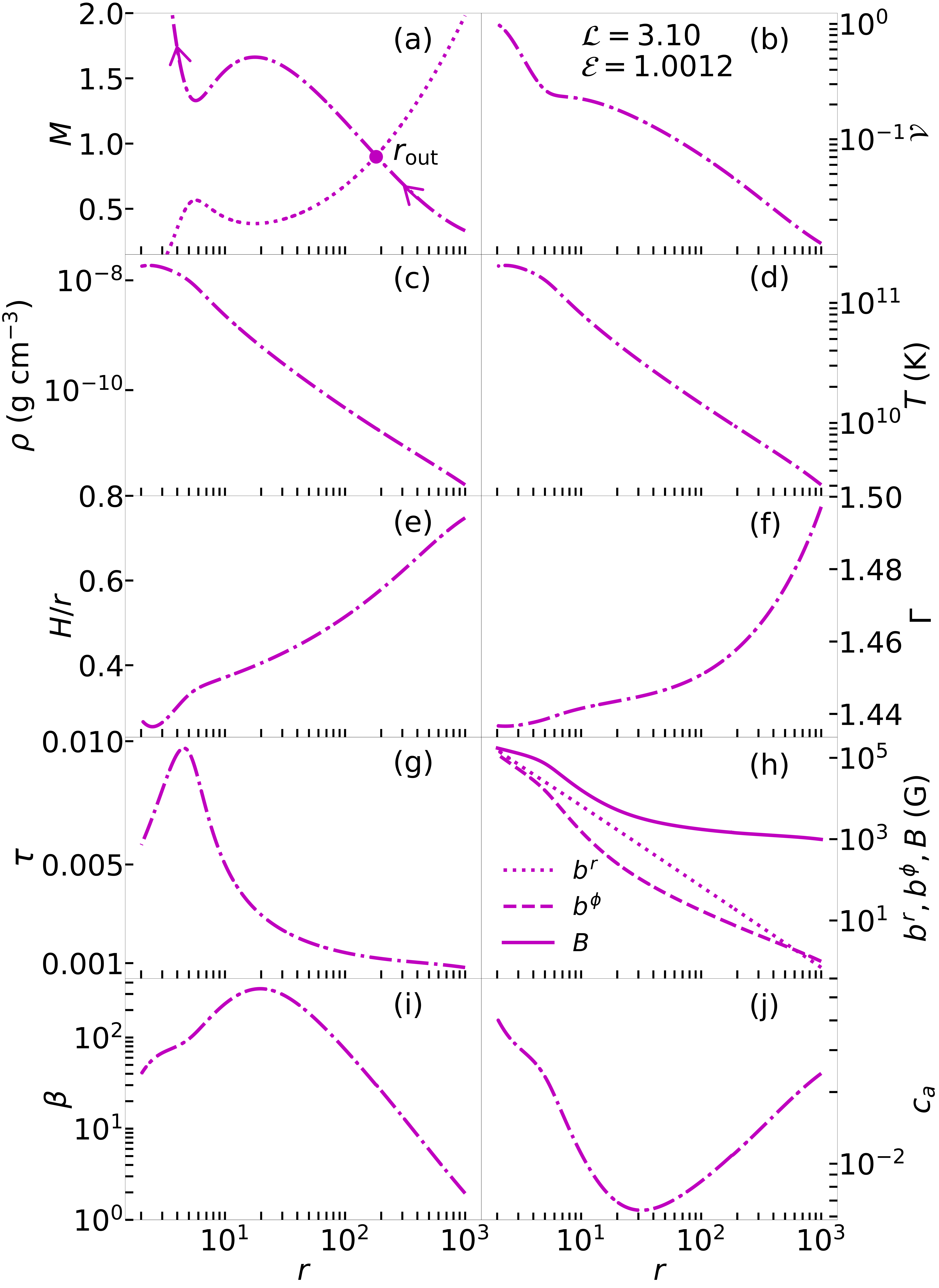}
	\caption{Same as Fig. \ref{EL_B_sol}, but the accretion solution eventually pass through the outer critical point $r_{\rm out}=181.465$ with ${\cal E}=1.0012$, ${\cal L}=3.10$, and $b^r_{\rm edge}=0.7$ G, respectively. See text for details. 
	}
	\label{M_r_out}
\end{figure}

For completeness, we continue to emphasize the importance of the GRMHD accretion flows that pass through the outer critical point, and study the primitive variables associated with the flow. For that we consider the accretion solution marked as `$S_4$' in Fig. \ref{M_r_in_out_Br}, and plot the profile of the corresponding flow variables, namely $M$, $v$, $\rho$, $T$, $H/r$, $\Gamma$, $\tau$, $B$, $b^r$, $b^\phi$, plasma $\beta$, and $c_a$  as function of radial distance in the respective panels (a)$-$(j) of Fig. \ref{M_r_out}. We observe that the accreting flow attains supersonic speed at a relatively larger radius ($r_{\rm out}$) in comparison to the accretion solutions possessing the inner critical points ($r_{\rm in}$) (see Fig. \ref{M_r_out}(a)). 
Because of this, the profiles of the primitive variables for `$S_4$' differ quantitatively from the solution `$S_1$' particularly at lower radii ($r \lesssim 10 r_g$), although their qualitative behaviour appear to be similar (see Fig. \ref{EL_B_sol}).  Nevertheless, it is noteworthy to mention that the accreting matter largely remains gas pressure dominated even at the near horizon limit ($r \lesssim 10 r_g$) although the strength of the magnetic fields reaches to $B \sim 0.75 \times 10^5$ G at the vicinity of the horizon (see Fig. \ref{M_r_out}(h)). This evidently signifies that the accretion disk presumably becomes magnetically more active for flows passing through $r_{\rm in}$ rather than $r_{\rm out}$ (see Fig. \ref{EL_B_sol}(h)).

\begin{figure}
	\includegraphics[width=\columnwidth]{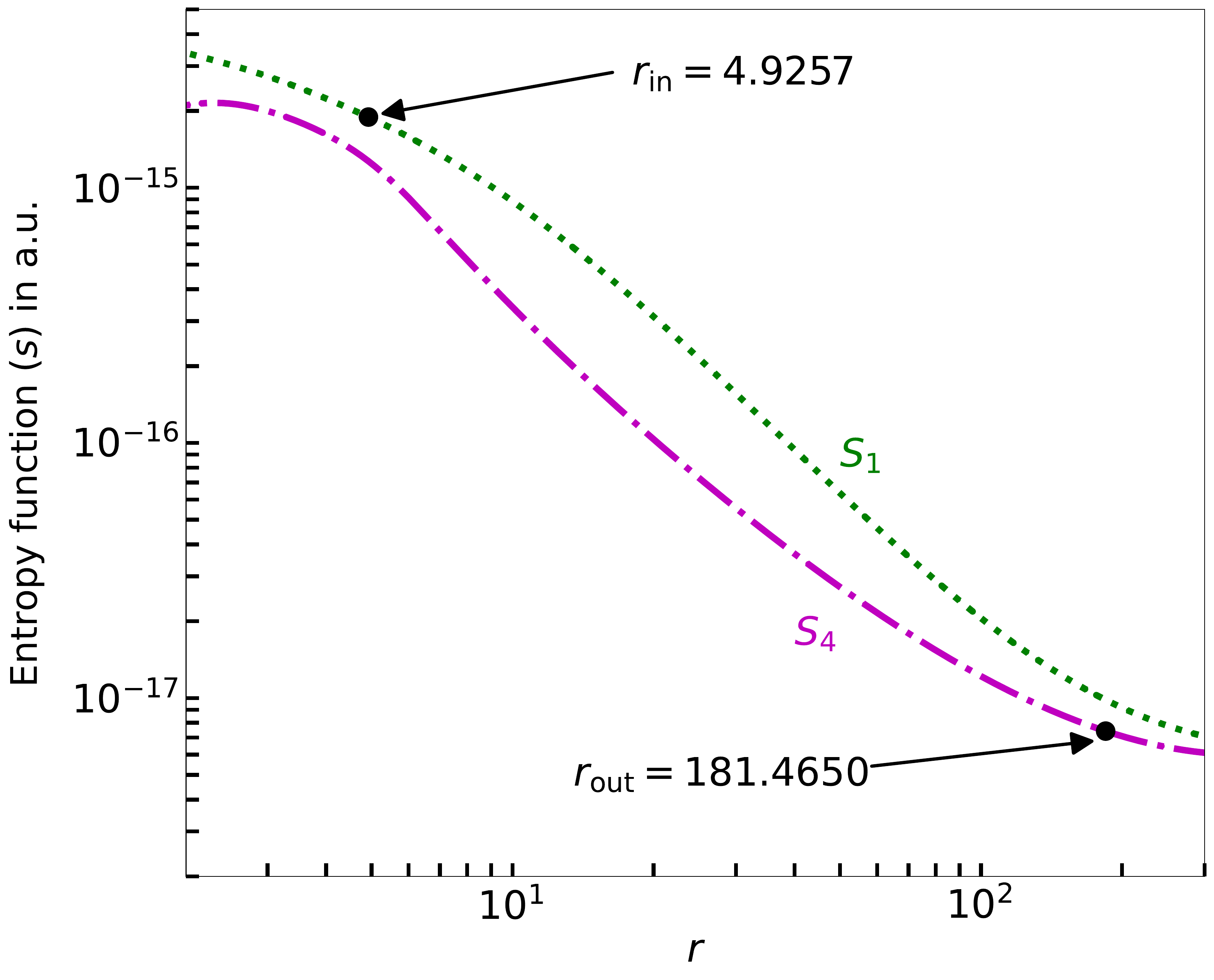}
	\caption{Variation of the specific entropy function ($s$) as a function of radial distance ($r$). Dotted and dot-dashed curves denote the results corresponding to the accretion solutions marked `$S_1$' and `$S_4$' in Fig. \ref{M_r_in_out_Br}. In the figure, inner critical point ($r_{\rm in}$) and outer critical point ($r_{\rm out}$) are marked. See text for details. 
	}
	\label{entropy0}
\end{figure}

	Next, in Fig. \ref{entropy0}, we compare the specific entropy function $s \propto p_{\rm tot}/\rho^{\Gamma -1}$ \citep{Das-etal2009,Porth-etal2017} corresponding to the accretion solutions passing through $r_{\rm in}$ and $r_{\rm out}$. While doing this, we consider solutions marked as `$S_1$' and `$S_4$' in Fig. \ref{M_r_in_out_Br} and depict the profile of $s$ corresponding to these solutions using dotted ($S_1$) and dot-dashed ($S_4$) curves. The inner and outer critical point locations are marked using the filled circles. We find that in both the cases, $s$ increases as the flow proceeds towards the black hole. This happens as a result of dissipation yielded in the differentially rotating magnetized flow around the black hole. What is more is that $s$ is seen to remain higher at all radii for solution passing through $r_{\rm in}$ compared to the solution possessing $r_{\rm out}$. A point worth mentioning here is that the global GRMHD accretion solution that changes its sonic character at $r_{\rm in}$ instead of $r_{\rm out}$ to become transonic, is perhaps thermodynamically favourable as it possess high entropy content, although their outer boundary conditions differ only by means of $b^r_{\rm edge}$ values.

	\section{conclusions}
	\label{Conclusion}
	
	In this paper, we investigate the global structure of a steady, magnetized, advective accretion flow around a black hole. Towards this, we self-consistently solve a set of governing equations \citep[and references therein]{Anile1990,Porth-etal2017} that regulate the dynamical structure of the MHD flow under the general relativistic framework and obtain the global transonic accretion solutions. Subsequently, we examine the properties of the accretion flow in terms of input parameters, namely, energy (${\cal E}$), angular momentum (${\cal L}$) and radial magnetic field ($b^r$), respectively. The findings of this work are summarized as follows.
	
	\begin{itemize}
		
		\item We obtain a complete set of global GRMHD accretion solutions around the black hole and find that accretion flow passes through either inner critical point ($r_{\rm in}$) or outer critical point ($r_{\rm out}$) before entering in to the black hole. We further notice that for a given (${\cal E}, {\cal L}$), when radial magnetic field at the outer edge of the disk is below a minimum value ($b^{r, \rm min}_{\rm edge}$), accretion flow possessing $r_{\rm in}$ changes its character and moves through $r_{\rm out}$ instead of $r_{\rm in}$ while approaching the black hole (see Fig. \ref{M_r_in_out_Br}). Similar findings are also observed when ${\cal L}$ or ${\cal E}$ are varied keeping other input parameters unchanged (see Figs. \ref{M_r_in_out_L} and \ref{M_r_in_out_E}).
		
		\item We observe that accretion flow remains mostly gas pressure dominated throughout the disk ($\beta > 10$) except at the near horizon limit $\lesssim 10 r_g$, where magnetic fields are seen to become considerably active ($\beta \sim 1$) (see Fig. \ref{BPBetaLambda_in}b). We also notice that the magnetic field strength is largely dominated by the toroidal field ($b^\phi b_\phi$) at all radii over the radial field ($b^r b_r$) except at the inner edge $r \lesssim 3r_g$ (see Fig. \ref{BPBetaLambda_in}c). We obtain the robust estimate of magnetic field strength over the entire length scale of the disk and observe that $B$ monotonically increases with the decrease of radial distance. For a solar mass ($M_{\rm BH} = M_{\odot}$) black hole, the magnetic field strength becomes very strong ($\sim 10^{6}$ Gauss) in the region close to the horizon (see Fig. \ref{BPBetaLambda_in}a). We also compute the viscosity parameter ($\alpha$) that governs the transport of specific angular momentum ($\lambda$) by means of Maxwell stress ($T^{\rm MAX}_{r \phi}$). We observe that unlike in standard disk \citep{Shakura-Sunyaev1973}, $\alpha$ varies with $r$ (see Fig. \ref{BPBetaLambda_in}f) and its profile agrees with the results from magnetically arrested disk (MAD) simulations \citep{Avara-etal2016}.
		
		\item We attempt to elucidate the radial profile of $\rho$, $p_{\rm gas}$, $p_{\rm mag}$, $b^r$ and $b^\phi$ by means of best fit power-law distribution. We find that all these flow variables can be ascertained satisfactorily as $\rho \propto r^{-(n+1/2)}$, $p_{\rm gas} \propto r^{-(n+7/6)}$, $p_{\rm mag} \propto r^{-(n+5/2)}$, $b^r \propto r^{-(n+1)}$, and $b^\phi \propto r^{-(n+9/5)}$, where $n\sim 1$. For pure accretion (no outflow), the density profile appears to be consistent with the results of \cite{Narayan-Yi1995,Blandford-Begelman2004} (see Fig. \ref{power-law-fit}). We further examine the correlation between $\alpha$ and plasma-$\beta$ that generally assumes the form of power law as $\alpha \propto \beta^{-q}$. For $6 r_g \lesssim r < 50 r_g$, we obtain $\alpha \propto \beta^{-2/5}$ which are in close agreement with results of the local shearing box simulation \citep{Salvesen-etal2016}. Interestingly, to the best of our knowledge, we find a new scaling relation yielding $\alpha \propto \beta^{-7/5}$ in the region ($2 r_g < r \lesssim 6 r_g$), where the disk is magnetically active (see Fig. \ref{alp_beta}). 

		\item It may be noted that depending on the input parameters, the accretion solution passing through $r_{\rm in}$ may not extend up to $r_{\rm edge}$ as it becomes closed at $r < r_{\rm edge}$ (see Fig. \ref{M_r_rin_br_vary}). In reality, solutions of this kind are potentially promising as they can be a part of global shocked accretion flow. Considering this, we identify the effective domain of three dimensional parameter space in (${\cal L}, {\cal E}, r_{\rm in}$) for a given set of ($b^r_{\rm in}, b^\phi_{\rm in}$) that admits closed GRMHD accretion solutions possessing $r_{\rm in}$. Generally, it appears that for	smaller ${\cal L}$, one requires higher $r_{\rm in}$ to obtain the closed solution and vice versa (see Fig. \ref{parasapce}).
	\end{itemize}
	
	With the above findings, we wish to emphasize that magnetic fields play pivotal role in regulating the structure as well as the dynamics of the GRMHD accretion flow around black hole. Overall, it is intriguing that the present formalism provides an insight of GRMHD accretion solution in the steady state limit and it would be useful in carrying out the state-of-the-art GRMHD simulation studies in higher dimensions which we plan to take up in future works.
	
	Finally, we wish to mention the limitations of the present work, as it is developed based on some approximations. We ignore the rotation of the black hole and also neglect mass loss from the disk. Further, we ignore the vertical component of the magnetic fields although it is expected to be relevant in launching jets and/or outflows \citep{Blandford-Payne1982,Blandford-Znajek1977,Koide-etal1998,Dihingia-etal2021}. In addition, we neglect the radiative cooling processes as well. Of course, the implementation of such issues is beyond the scope of this paper, however, we argue that the overall findings of the present analysis will remain qualitatively intact.
	
	We also state that in this work, we adopt ideal GRMHD approximation as an introductory approach for the purpose of simplicity, although the works involving complex non-ideal MHD approximations are more suitable which we plan to consider for future endeavour.

\section*{Data availability statement}

The data underlying this article will be available with reasonable request.
	
	\section*{Acknowledgements}
	
	We thank the anonymous reviewers for constructive comments and useful suggestions that help to improve the quality of the paper. SM acknowledges Prime Minister's Research Fellowship (PMRF), Government of India for financial support. IKD thanks the financial support from Max Planck partner group award at Indian Institute Technology of Indore (MPG-01). This work was supported by the Science and Engineering Research Board (SERB) of India through grant MTR/2020/000331. 

	\input{ms_01.bbl}
	
	\appendix

	\onecolumn
	\section{Calculation of Wind Equation}\label{Wind}
	
	Using equation (18) in equations (11,12,13,14, and 16), we obtain,  
		\begin{equation}
			\label{28}
			R_0 + R_v \frac{dv}{dr} + R_\Theta \frac{d\Theta}{dr} + R_\lambda \frac{d\lambda}{dr} + R_{b_r} \frac{db^r}{dr} + R_{b_\phi} \frac{db^\phi}{dr} =0,
		\end{equation}

		\begin{equation}
			\label{29}
			\mathcal{L}_0 + \mathcal{L}_v \frac{dv}{dr} + \mathcal{L}_\Theta \frac{d\Theta}{dr} + \mathcal{L}_\lambda \frac{d\lambda}{dr} + \mathcal{L}_{b_r} \frac{db^r}{dr} + \mathcal{L}_{b_\phi} \frac{db^\phi}{dr}=0,
		\end{equation}

		\begin{equation}
			\label{30}
			\mathcal{E}_0 + \mathcal{E}_v \frac{dv}{dr} + \mathcal{E}_\Theta \frac{d\Theta}{dr} + \mathcal{E}_\lambda \frac{d\lambda}{dr} + \mathcal{E}_{b_r} \frac{db^r}{dr} + \mathcal{E}_{b_\phi} \frac{db^\phi}{dr}=0, 
		\end{equation} 

		\begin{equation}
			\label{31}
			\mathcal{B}_{r_0} + \mathcal{B}_{r_v} \frac{dv}{dr} + \mathcal{B}_{r_\Theta} \frac{d\Theta}{dr} + \mathcal{B}_{r_\lambda} \frac{d\lambda}{dr} + \mathcal{B}_{r_{b_r}} \frac{db^r}{dr} + \mathcal{B}_{r_{b_\phi}} \frac{db^\phi}{dr}=0, 
		\end{equation} 

		\begin{equation}
			\label{32}
			\mathcal{B}_{\phi_0} + \mathcal{B}_{\phi_v} \frac{dv}{dr} + \mathcal{B}_{\phi_\Theta} \frac{d\Theta}{dr} + \mathcal{B}_{\phi_\lambda} \frac{d\lambda}{dr} + \mathcal{B}_{\phi_{b_r}} \frac{db^r}{dr} + \mathcal{B}_{\phi_{b_\phi}} \frac{db^\phi}{dr}=0.
		\end{equation} 

	The coefficients of the equations (A1-A5) take the form,\\
	$$R_0 = \big(R_i + \mathcal{A} R_2\big)/\rho h_{\rm tot} \mathcal{A}, \mathcal{A}=(g^{rr}+u^{r}u^r), R_2 = -\frac{g_{t\phi} u_{r0} b^{r}b^{\phi}}{u_t} + \frac{\mathcal{F}_1 \Theta  \rho }{\mathcal{F} \tau }-\frac{3 \Theta  \rho }{r \tau} + h_{\rm tot} \rho  u^r u^r{}_0 + R_1,$$ 
	$$R_1= \frac{ g_{tt} u_r u_{t0} b^{r}b^{t}}{u_t^2} -\frac{ g_{tt} u_{r0} b^{r}b^{t}}{u_t} + \frac{g_{t\phi} u_r u_{t0} b^{r}b^{\phi }}{u_t^2} -\frac{\Theta  \rho  \Delta '}{\Delta  \tau } + \frac{1}{2} b^{\phi ^2} g_{\phi \phi}' + b^{t}b^{\phi } g_{t\phi}'+ \frac{1}{2} b^{t^2} g_{tt}', \Delta'=\frac{\partial \Delta}{\partial r}$$ 
	$$\mathcal{F}'=\frac{d\mathcal{F}}{dr}=\mathcal{F}_1 + \mathcal{F}_2 \frac{d\lambda}{dr},
	\mathcal{F}_1 = \frac{\partial\mathcal{F}}{\partial r}, 
	\mathcal{F}_2 = \frac{\partial\mathcal{F}}{\partial \lambda}, u_{\mu}'= u_{\mu_0} + u_{\mu_v} v' + u_{\mu_\lambda} \lambda',	u^{\mu'}= u^{\mu}_0 + u^{\mu}_v v' + u^{\mu}_\lambda \lambda'; \mu \equiv (t,r,0,\phi),$$
	$$u^{\mu}_0=\frac{\partial u^\mu}{\partial r}, u^\mu_v=\frac{\partial u^\mu}{\partial v}, u^\mu_\lambda=\frac{\partial u^\mu}{\partial \lambda}, u_{\mu_0}=\frac{\partial u_\mu}{\partial r}, u_{\mu_v}=\frac{\partial u_\mu}{\partial v}, u_{\mu_\lambda}=\frac{\partial u_\mu}{\partial \lambda}.$$	
	$$R_i = -2 u^r b^{r}b^{\phi } \left(g_{t\phi} u^t+g_{\phi \phi } u^{\phi }\right) \left(\frac{1}{2} g^{\phi \phi } g_{\phi \phi}'+\frac{1}{2} g^{t\phi}	g_{t\phi}'\right)  -b^{r^2} \left(\frac{1}{2} g^{\phi \phi } g _{\phi \phi}'+\frac{1}{2} g^{t\phi} g_{t\phi}'\right) + R_h + R_g,$$
	$$R_h = \rho h_{\rm tot} u^{\phi } \left(-\frac{1}{2} g^{rr} u^t	g_{t\phi}'-\frac{1}{2} g^{rr} u^{\phi}g_{\phi\phi}'\right), R_g=-b^{\phi } g_{rr} u^{r^^2} \left(-\frac{1}{2} b^{\phi }
	g^{rr} g_{\phi \phi}'-\frac{1}{2} b^t g^{rr} g_{t\phi}'\right) + R_f,$$
	$$R_f = -2 u^r b^{r}b^{\phi } \left(g_{tt} u^t+g_{t\phi} u^{\phi }\right) \left(\frac{1}{2} g^{t\phi} g_{\phi \phi }'+\frac{1}{2} g^{tt} g_{t\phi}'\right) + R_e,  R_e = -2 u^r b^{r}b^{t} \left(g_{t\phi} u^t+g_{\phi \phi } u^{\phi
	}\right) \left(\frac{1}{2} g^{\phi \phi } g_{t\phi}'+\frac{1}{2} g^{t\phi} g_{tt}'\right) + R_d,$$ 
	$$R_d = \frac{1}{2} b^{\phi ^2} g^{rr} g_{\phi \phi}'-\frac{1}{2} b^{r^2} g^{\theta \theta } g_{\theta \theta}' + \rho h_{\rm tot}  u^t \left(-\frac{1}{2} g^{rr} u^t g_{tt}'-\frac{1}{2} g^{rr} u^{\phi } g_{t\phi}'\right) -b^t g_{rr} u^{r^2} \left(-\frac{1}{2} b^{\phi} g^{rr} g_{t\phi}'-\frac{1}{2} b^t g^{rr}g_{tt}'\right)+R_c,$$ 
	$$R_c=-2 u^r b^{r}b^{t} \left(g_{tt} u^t+g_{t\phi} u^{\phi
	}\right) \left(\frac{1}{2} g^{t\phi} g_{t\phi}'+\frac{1}{2} g^{tt} g_{tt}'\right)-b^{r^2}\left(\frac{1}{2} g^{t\phi} g_{t\phi}'+\frac{1}{2} g^{tt} g_{tt}'\right) - b^{r^2} g^{rr} g_{rr}'-\frac{1}{2}
	b^{r^2} g^{rr} g_{rr} u^{r^2} g_{rr}'+R_b,$$  $$R_b=\frac{1}{2} b^{t^2} g^{rr} g_{tt}'+\frac{1}{2}
	\rho h_{\rm tot}  g^{rr} u^{2 r} g_{rr}' + g^{rr} b^{t}b^{\phi} g_{t\phi}' -\frac{b^{r^2} u^r u_r u_{t_0} \left(g_{tt} u^t+g_{t\phi} u^{\phi }\right)}{u_t^2} + R_a, R_a = h_{\rm tot} \rho u^r u^r{}_0 + \frac{b^{r^2} u^r u_{r_0} \left(g_{tt} u^t+g_{t\phi} u^{\phi }\right)}{u_t},$$
	$$R_{v}=\frac{R_{v 0}+\mathcal{A} R_{v 1}}{\mathcal{A}\rho h_{\rm tot}},   R_{v0}=\frac{b^{r^2} u^r u_{rv} \left(g_{tt} u^t+g_{t\phi} u^{\phi }\right)}{u_t}-\frac{b^{r^2} u^r
		u_r u_{tv} \left(g_{tt} u^t+g_{t\phi} u^{\phi }\right)}{u_t^2},$$  $$R_{v1}=\frac{ g_{t\phi} u_r u_{t\lambda} b^{r}b^{\phi}}{u_t^2}+\frac{ g_{tt} u_r u_{t_\lambda}b^{r}b^{t}}{u_t^2}+ b^{\phi ^2} g_{t\phi}+\frac{ \mathcal{F}_2
		\Theta  \rho }{\mathcal{F} \tau }+g_{tt} b^{t}b^{\phi }, R_\Theta=\frac{1}{h_{\text{tot}} \tau }, R_{\lambda}=\frac{R_{\lambda 0}+R_{\lambda 1}- u^r b^{r}b^{\phi } \left(g_{tt}u^t+g_{t\phi} u^{\phi }\right)}{\mathcal{A}\rho h_{\rm tot}},$$
	$$R_{\lambda 0}=\mathcal{A}\bigg(\frac{g_{t\phi} u_r u_{t\lambda} b^{r}b^{\phi}}{u_t^2}+\frac{ \mathcal{F}_2 \Theta  \rho }{\mathcal{F} \tau } + \frac{g_{tt} u_r u_{t\lambda} b^{r}b^{t}}{u_t^2}+b^{\phi ^2} g_{t\phi}\bigg), R_{\lambda 1}=\mathcal{A} g_{tt} b^{t}b^{\phi}-\frac{b^{r^2} u^r u_r u_{t\lambda} \left(g_{tt} u^t+g_{t\phi} u^{\phi }\right)}{u_t^2},$$
	$$R_{b_r}=\bigg(R_{b_{r 0}} + \frac{b^r u^r u_r \left(g_{tt} u^t+g_{t\phi}
		u^{\phi }\right)}{u_t}\bigg)/\rho h_{\rm tot} \mathcal{A},  R_{b_{r 0}}=\mathcal{A}(g_{rr} b^r - \frac{g_{t\phi} u_rb^\phi}{u_t} - \frac{g_{tt} u_r b^t}{u_t}) -2 b^r - g_{rr} b^r u^{r^2},$$
	$$R_{b_\phi} = \bigg(R_{b_{\phi 0}}- b^r u^r \lambda (g_{tt} u^t + g_{t\phi} u^\phi) - b^r u^r (g_{t\phi} u^t + g_{\phi\phi} u^\phi)\bigg)/\rho h_{\rm tot} \mathcal{A}, R_{b_{\phi 0}}= \mathcal{A} \big(g_{\phi\phi} b^{\phi} + g_{t\phi} \lambda b^{\phi} + g_{t\phi} b^t + g_{tt} \lambda b^t\big),$$
	$$\mathcal{L}_{0}=\frac{B^2 \Delta' u_{\phi }}{2 \Delta  \rho }-\frac{b^r \Delta'  \left(b^{\phi} g_{\phi \phi }+b^t g_{t\phi}\right)}{2 \Delta \rho u^{r}} -\frac{b^{r^2} g_{t\phi} u_r u_{t_0}}{\rho u^{r} u_t^2} -\frac{2 g_{t\phi} u_{r_0} u_{\phi } b^{r}b^{\phi }}{\rho u_t}+\frac{b^{r^2} g_{t\phi}  u_{r_0}}{\rho u^{r} u_t}-\frac{2 g_{tt} u_{r_0} u_{\phi } b^{r}b^{t}}{\rho  u_t} + \mathcal{L}_{01},$$ 
	$$\mathcal{L}_{01}=\frac{2 g_{t\phi} u_r u_{t_0} u_{\phi } b^{r}b^{\phi }}{\rho u_t^2} + \frac{2 g_{tt} u_r u_{t_0} u_{\phi } b^{r}b^{t}}{\rho  u_t^2}+\frac{b^{\phi ^2} u_{\phi } g_{\phi \phi}'}{\rho} -\frac{3 b^r\left(b^{\phi } g_{\phi \phi }+b^t g_{t\phi}\right)}{2 \rho u^r r} + \frac{3 B^2 u_{\phi }}{2\rho  r}-\frac{B^2 \mathcal{F}_1 u_{\phi }}{2 \mathcal{F} \rho } + \mathcal{L}_{02},$$
	$$\mathcal{L}_{02}=\frac{\mathcal{F}_1 b^r \left(b^{\phi } g_{\phi \phi }+b^t g_{t\phi}\right)}{2
		\mathcal{F} \rho u^r}+\frac{b^r u^r{}_0 \left(b^{\phi } g_{\phi \phi }+b^t g_{t\phi}\right)}{\rho u^{r^2}}+\frac{b^{r^2} u_{\phi } g_{rr}'}{\rho }+u_{\phi 0} h_{\rm tot} + \frac{2 u_\phi b^{t}b^{\phi } g_{t\phi}'}{\text{$\rho $}}-\frac{ b^{r}b^{\phi } g_{\phi \phi}'}{\rho u^r}+\frac{b^{t^2} u_{\phi } g_{tt}'}{\rho }-\frac{ b^{r}b^{t} g_{t\phi}'}{\rho u^r},$$
	$$\mathcal{L}_{v}=-\frac{2 g_{t\phi} u_{rv} u_{\phi } b^{r}b^{\phi }}{\rho u_t}+\frac{b^{r^2} g_{t\phi}  u_{rv}}{\rho u^{r} u_t}-\frac{b^{r^2} g_{t\phi} u_r u_{tv}}{\rho u^{r} u_t^2} + \frac{2 g_{t\phi} u_r u_{tv} u_{\phi } b^{r}b^{\phi }}{\rho u_t^2}-\frac{2 g_{tt} u_{rv} u_{\phi } b^{r}b^{t}}{\rho  u_t} + u_{\phi v} h_{\rm tot}+\frac{B^2 u_{\phi} \left(v^2 \gamma _v^2+1\right)}{\rho  v} + \mathcal{L}_{v1},$$
	$$\mathcal{L}_{v1}=\frac{b^r u^r{}_v \left(b^{\phi } g_{\phi \phi }+b^t g_{t\phi}\right)}{\rho u^{r^2}}-\frac{b^r \left(v^2 \gamma _v^2+1\right) \left(b^{\phi }
		g_{\phi \phi }+b^t g_{t\phi}\right)}{\rho u^{r} v}, \mathcal{L}_{\Theta}=-\frac{b^r \left(b^{\phi } g_{\phi \phi }+b^t g_{t\phi}\right)}{2\Theta  \rho u^r}+\frac{u_{\phi }}{\tau} \frac{df}{d\Theta}+\frac{2 u_{\phi}}{\tau } + \frac{B^2 u_{\phi }}{2 \Theta  \rho },$$
	$$\mathcal{L}_{\lambda}=\frac{2 g_{tt} u_{\phi} b^{t}b^{\phi}}{\text{$\rho $}}+\frac{\mathcal{F}_2
		b^r \left(b^{\phi } g_{\phi \phi }+b^t g_{t\phi}\right)}{2 \mathcal{F} \rho u^{r}} + \frac{2 g_{t\phi} u_r u_{t\lambda} u_{\phi } b^{r}b^{\phi }}{\rho u_t^2}-\frac{g_{t\phi} b^{r}b^{\phi }}{\rho u^{r}}-\frac{b^{r^2} g_{t\phi} u_r u_{t\lambda }}{\rho u^{r} u_t^2}-\frac{B^2 \mathcal{F}_2 u_{\phi }}{2 \mathcal{F} \rho } + \mathcal{L}_{\lambda1},$$
	$$\mathcal{L}_{\lambda1}=\frac{2 g_{tt} u_r u_{t\lambda} u_{\phi } b^{r}b^{t}}{\rho u_t^2}+\frac{2 b^{\phi ^2} g_{t\phi} u_{\phi }}{\rho } +u_{\phi \lambda }h_{\rm tot}, 	\mathcal{L}_{b_r}=-\frac{2 b^{\phi } g_{t\phi} u_r u_{\phi }}{\rho  u_t}+\frac{2 b^r g_{rr} u_{\phi }}{\rho }+\frac{b^r g_{t\phi} u_r}{\rho u^{r} u_t} -\frac{2 b^t g_{tt} u_r u_{\phi }}{\rho  u_t}-\frac{\left(b^{\phi } g_{\phi \phi }+b^t g_{t\phi}\right)}{\rho u^r},$$
	$$\mathcal{L}_{b_{\phi}}=\frac{2 \lambda  b^{\phi } g_{t\phi} u_{\phi } + 2 b^{\phi } g_{\phi \phi } u_{\phi }}{\rho }-\frac{\lambda  b^r g_{t\phi} }{\rho u^{r}}-\frac{b^r g_{\phi \phi }}{\rho u^{r}}+\frac{2 \lambda  b^t g_{tt} u_{\phi }}{\rho } +  \frac{2 b^t g_{t\phi} u_{\phi }}{\rho},$$
	$$\mathcal{E}_{0}=\frac{b^r \Delta ' \left(b^{\phi } g_{t\phi}+b^t g_{tt}\right)}{2 \Delta  \rho u^{r}}-\frac{\Delta ' u_t B^2}{2 \Delta  \rho }+\frac{ b^{r}b^{\phi } g_{t\phi}'}{u^{r}\rho }-\frac{b^{t^2} u_t g_{tt}'}{\rho}-\frac{b^{r^2} g_{tt}  u_{r_0}}{\rho  u_t u^{r}} + \frac{b^{r^2} g_{tt} u_r u_{t_0}}{\rho u_t^2 u^{r}} + \frac{2 g_{t\phi} u_{r_0} b^{r}b^{\phi }}{\rho }-\frac{2 g_{t\phi} u_r u_{t_0} b^{r}b^{\phi }}{\rho  u_t} + \mathcal{E}_{01},$$
	$$\mathcal{E}_{01}=\frac{2 g_{tt} u_{r_0} b^{r}b^{t}}{\rho }-\frac{2 g_{tt} u_r u_{t_0} b^{r}b^{t}}{\rho  u_t}+\frac{3 b^r \left(b^{\phi } g_{t\phi}+b^t g_{tt}\right)}{2 u^r \rho  r}-\frac{b^{\phi ^2} u_t g_{\phi \phi}'}{\rho} -\frac{\mathcal{F}_1 b^r \left(b^{\phi } g_{t\phi}+b^t g_{tt}\right)}{2 u^{r} \mathcal{F} \rho } -\frac{b^r  u^r{}_0 \left(b^{\phi } g_{t\phi}+b^t g_{tt}\right)}{u^{r^2} \rho} -\frac{2 u_t b^{t}b^{\phi } g_{t\phi}'}{\text{$\rho $}} + \mathcal{E}_{02},$$
	$$\mathcal{E}_{02}=-\frac{3 u_t B^2}{2 \rho  r} + \frac{B^2 \mathcal{F}_1 u_t}{2 \mathcal{F} \rho }-h_{\text{tot}} u_{t_0}+\frac{ b^{r}b^{t} g_{tt}'}{ u^{r} \rho }-\frac{b^{r^2} u_t g_{rr}'}{\rho}+\frac{\mathcal{F}_1 u_t \left(2 g_{t\phi} b^{t}b^{\phi }+b^{r^2} g_{rr}+b^{t^2} g_{tt}+b^{\phi ^2} g_{\phi \phi }\right)}{2 \mathcal{F} \rho },$$
	$$\mathcal{E}_{v}=\frac{2 g_{t\phi} u_{rv} b^{r}b^{\phi }}{\rho }-\frac{b^{r^2} g_{tt} u_{rv}}{\rho u^{r}  u_t}+\frac{b^{r^2} g_{tt} u_r u_{tv}}{\rho u^{r} u_t^2}+\frac{2 g_{tt} u_{rv} b^{r}b^{t}}{\rho}-\frac{2 g_{t\phi} u_r u_{tv} b^{r}b^{\phi }}{\rho u_t}-\frac{2 g_{tt} u_r u_{tv} b^{r}b^{t}}{\rho  u_t}-\frac{b^r u^r{}_v \left(b^{\phi } g_{t\phi}+b^t g_{tt}\right)}{\rho u^{r^2}} + \mathcal{E}_{v1},$$
	$$\mathcal{E}_{v1}=\frac{b^r \left(v^2 \gamma _v^2+1\right) \left(b^{\phi } g_{t\phi}+b^t g_{tt}\right)}{\rho u^{r} v}-\frac{B^2 u_t \left(v^2 \gamma _v^2+1\right)}{\rho  v} -h_{\text{tot}} u_{tv}, \mathcal{E}_{\Theta}=\frac{b^r \left(b^{\phi } g_{t\phi}+b^t g_{tt}\right)}{2 \rho \Theta u^{r}}-\frac{B^2 u_t}{2 \rho  \Theta }-\frac{u_t}{\tau } \frac{df}{d\Theta}-\frac{2 u_t}{\tau },$$
	$$\mathcal{E}_{\lambda}=-\frac{2 g_{t\phi} u_r u_{t\lambda} b^{r}b^{\phi }}{\rho u_t}+\frac{g_{tt} b^{r}b^{\phi }}{\rho  u^{r}}+\frac{b^{r^2} g_{tt} u_r u_{t\lambda}}{\rho u^{r} u_t^2}-\frac{2 b^{\phi^2} g_{t\phi} u_t}{\rho }-\frac{2 g_{tt} u_r u_{t\lambda} b^{r}b^{t}}{\rho  u_t} -\frac{2 g_{tt} u_{t} b^{t}b^{\phi }}{\rho}-\frac{\mathcal{F}_2 b^r \left(b^{\phi } g_{t\phi}+b^t g_{tt}\right)}{2 \mathcal{F} u^{r} \rho } + \frac{B^2 \mathcal{F}_2 u_t}{2 \mathcal{F} \rho } - h_{\text{tot}} u_{t\lambda},$$
	$$\mathcal{E}_{b_{r}}=\frac{2 b^{\phi } g_{t\phi} u_r}{\rho }+\frac{2 b^t g_{tt} u_r}{\rho}-\frac{2 b^r g_{rr} u_t}{\rho }-\frac{b^r g_{tt} u_r }{\rho u^{r} u_t} + \frac{ \left(b^{\phi } g_{t\phi}+b^t g_{tt}\right)}{\rho u^{r}},$$
	$$\mathcal{ E}_{b_{\phi}}=\bigg(-2 \lambda  b^{\phi } g_{t\phi} u_t+2 b^{\phi } g_{\phi \phi } u_t+2 \lambda  b^t g_{tt} u_t+2 b^t g_{t\phi} u_t +\frac{\lambda  b^r g_{tt} }{u^{r}}+\frac{b^r g_{t\phi}}{u^{r}} \bigg)/\rho,$$
	$$B_{r_0}=r\big( B_{r_{00}} - B_{r_{01}} - B_{r_{02}}\big), \mathcal{B}_{r_{00}}=2 \lambda  b^{\phi } u^r+\lambda r b^{\phi }	u^r{}_0-\frac{r b^r u^r u_{r_0}}{u_t}+\frac{r b^r u^r u_r u_{t_0}}{u_t^2},  B_{r_{01}} = 2 b^r u^t-r b^r u^t{}_0,$$  
	$$B_{r_{02}} = \frac{r b^r u_r u^r{}_0}{u_t}-\frac{2 b^r u^r u_r}{u_t}, \mathcal{B}_{r_{v}}=r \bigg( \lambda  r b^{\phi } u^r{}_v-\frac{r b^r u^r u_{rv}}{u_t}+\frac{r b^r u^r u_r u_{tv}}{u_t^2}-r b^r u^t{}_v -\frac{r b^r u_r u^r{}_v}{u_t}\bigg),
	\mathcal{B}_{r_{b_\phi}}=\lambda  r^2 u^r,$$ 
	$$\mathcal{B}_{r_{b_r}}=r \left(-r u^t-\frac{r u_r u^r}{u_t}\right),\hspace{0.2cm}\mathcal{B}_{r_\Theta}=0,\hspace{0.15cm}
	\mathcal{B}_{r_\lambda}=r \left(r b^{\phi } u^r+\frac{r b^r u^r u_r u_{t\lambda}}{u_t^2}-r b^r u^t{}_{\lambda }\right),$$

	$$\mathcal{B}_{\phi_0}=r^2 b^{\phi } u^r{}_0-r^2 b^r u^{\phi }{}_0+2 r \left(b^{\phi }
	u^r-b^r u^{\phi }\right), \hspace{0.2cm}
	\mathcal{B}_{\phi_v}=r^2 b^{\phi } u^r{}_v-r^2 b^r u^{\phi }{}_v, \hspace{0.2cm}\mathcal{B}_{\phi_\Theta}=0, $$
	$$\mathcal{B}_{\phi_{\lambda}}=-r^2 -b^r u^{\phi }{}_{\lambda }, \hspace{0.15cm}
	\mathcal{B}_{\phi_{b_r}}=-r^2 u^{\phi },\hspace{0.2cm} \mathcal{B}_{\phi_{b_\phi}}=r^2 u^r.$$
	
	With the help of equations (A1)-(A5), the wind equations is expressed as,
	\begin{equation}
		\frac{dv}{dr} = \frac{\mathcal{N}(r,v,\Theta,\lambda,b^r, b^\phi)}{\mathcal{D}(r,v,\Theta,\lambda,b^r, b^\phi)}
	\end{equation}
	\par 
	where, 
	\begin{equation}
		\label{38}
		\mathcal{N}(r,v,\Theta,\lambda,b^r, b^\phi) = -\bigg( R_0 + b^r_{11} R_{b_r} + R_\Theta \Theta_{11} + R_\lambda \lambda_{11} + R_{b_\phi} b^\phi_{11} \bigg)
	\end{equation}
	and 
	\begin{equation}
		\label{39}
		\mathcal{D}(r,v,\Theta,\lambda,b^r, b^\phi) = R_v + b^r_{12} R_{b_r}+ R_\Theta \Theta_{12} + R_\lambda \lambda_{12} + R_{b_\phi} b^\phi_{12}. 
	\end{equation}
	
	Here, are the remaining coefficients of above equations (A7, A8) as follows,
	\begin{flalign*}
		\Theta_{11} = -\frac{\Theta_{c0}}{\mathbb{D}}, \Theta_{12}=-\frac{\Theta_{v0}}{\mathbb{D}}, \lambda_{11} = \frac{\lambda_{c0}}{\mathbb{D}}, \lambda_{12} = \frac{\lambda_{v0}}{\mathbb{D}}, b^r_{11} =- \frac{b^r_{c0}}{\mathbb{D}}, b^r_{12} =- \frac{b^r_{v0}}{\mathbb{D}}, b^\phi_{11} =- \frac{b^\phi_{c0}}{\mathbb{D}}, b^\phi_{12} =- \frac{b^\phi_{v0}}{\mathbb{D}},
	\end{flalign*}
	where,
	\begin{equation*}
		\begin{aligned}
			&\mathbb{D}=\mathcal{B}_{r_{b_\phi}} \left(\mathcal{L}_{\lambda } \mathcal{E}_{b_r} \mathcal{B}_{\phi _{\Theta }}-\mathcal{E}_{\lambda } \mathcal{L}_{b_r}
			\mathcal{B}_{\phi _{\Theta }}+\mathcal{B}_{\phi _{\lambda }} \left(\mathcal{E}_{\Theta } \mathcal{L}_{b_r}-\mathcal{L}_{\Theta }
			\mathcal{E}_{b_r}\right)+\mathcal{E}_{\lambda } \mathcal{L}_{\Theta } \mathcal{B}_{\phi_{b_r}}-\mathcal{E}_{\Theta } \mathcal{L}_{\lambda }
			\mathcal{B}_{\phi_{b_r}}\right)+\mathcal{E}_{\lambda } \mathcal{L}_{\Theta } \left(-\mathcal{B}_{r_{b_r}}\right) \mathcal{B}_{\phi_{b_\phi}}+\mathcal{E}_{\Theta } \mathcal{L}_{\lambda } \mathcal{B}_{r_{b_r}} \mathcal{B}_{\phi_{b_\phi}}\\&+\mathcal{L}_{b_{\phi }} \left(\mathcal{E}_{b_r} \mathcal{B}_{r_{\Theta }} \mathcal{B}_{\phi _{\lambda
			}}+\mathcal{E}_{\Theta } \left(-\mathcal{B}_{r_{b_r}}\right) \mathcal{B}_{\phi _{\lambda }}+\mathcal{E}_{\lambda } \mathcal{B}_{r_{b_r}}
			\mathcal{B}_{\phi _{\Theta }}-\mathcal{E}_{\lambda } \mathcal{B}_{r_{\Theta }} \mathcal{B}_{\phi_{b_r}}\right)+\mathcal{L}_{\Theta }
			\mathcal{E}_{b_{\phi }} \mathcal{B}_{r_{b_r}} \mathcal{B}_{\phi _{\lambda }}-\mathcal{L}_{\lambda } \mathcal{E}_{b_{\phi }}
			\mathcal{B}_{r_{b_r}} \mathcal{B}_{\phi _{\Theta }}-\mathcal{L}_{\lambda } \mathcal{E}_{b_r} \mathcal{B}_{r_{\Theta }}
			\mathcal{B}_{\phi_{b_\phi}}\\&+\mathcal{L}_{\lambda } \mathcal{E}_{b_{\phi }} \mathcal{B}_{r_{\Theta }} \mathcal{B}_{\phi
				{b_r}}+\mathcal{E}_{\lambda } \mathcal{L}_{b_r} \mathcal{B}_{r_{\Theta }} \mathcal{B}_{\phi_{b_\phi}}+\mathcal{B}_{r_{\lambda }}
			\left(\mathcal{L}_{\Theta } \mathcal{E}_{b_r} \mathcal{B}_{\phi_{b_\phi}}-\mathcal{L}_{\Theta } \mathcal{E}_{b_{\phi }}
			\mathcal{B}_{\phi_{b_r}}-\mathcal{E}_{\Theta } \mathcal{L}_{b_r} \mathcal{B}_{\phi_{b_\phi}}+\mathcal{E}_{\Theta }
			\mathcal{L}_{b_{\phi }} \mathcal{B}_{\phi_{b_r}}+\mathcal{L}_{b_r} \mathcal{E}_{b_{\phi }} \mathcal{B}_{\phi _{\Theta }}-\mathcal{E}_{b_r}
			\mathcal{L}_{b_{\phi }} \mathcal{B}_{\phi _{\Theta }}\right)\\&-\mathcal{L}_{b_r} \mathcal{E}_{b_{\phi }} \mathcal{B}_{r_{\Theta }} \mathcal{B}_{\phi
				_{\lambda }},
		\end{aligned}
	\end{equation*}
	
	\begin{equation*}
		\begin{aligned} 
			&\Theta_{c0}=\mathcal{B}_{r_{b_\phi}} \left(\mathcal{B}_{\phi _0}
			\mathcal{L}_{\lambda } \mathcal{E}_{b_r}-\mathcal{B}_{\phi _0}
			\mathcal{E}_{\lambda } \mathcal{L}_{b_r}+\mathcal{B}_{\phi
				_{\lambda }} \left(\mathcal{E}_0 \mathcal{L}_{b_r}-\mathcal{L}_0
			\mathcal{E}_{b_r}\right)+\mathcal{L}_0 \mathcal{E}_{\lambda }
			\mathcal{B}_{\phi_{b_r}}-\mathcal{E}_0 \mathcal{L}_{\lambda
			} \mathcal{B}_{\phi_{b_r}}\right)+\mathcal{L}_0 \mathcal{E}_{\lambda }
			\left(-\mathcal{B}_{r_{b_r}}\right) \mathcal{B}_{\phi_{b_\phi}}+\mathcal{E}_0 \mathcal{L}_{\lambda }
			\mathcal{B}_{r_{b_r}} \mathcal{B}_{\phi_{b_\phi}}\\&+\mathcal{L}_{b_{\phi }}
			\left(\mathcal{B}_{r_0} \mathcal{E}_{b_r} \mathcal{B}_{\phi
				_{\lambda }}+\mathcal{E}_0 \left(-\mathcal{B}_{r_{b_r}}\right)
			\mathcal{B}_{\phi _{\lambda }}+\mathcal{B}_{\phi _0}
			\mathcal{E}_{\lambda } \mathcal{B}_{r_{b_r}}-\mathcal{B}_{r_0}
			\mathcal{E}_{\lambda } \mathcal{B}_{\phi
				{b_r}}\right)+\mathcal{L}_0 \mathcal{E}_{b_{\phi }}
			\mathcal{B}_{r_{b_r}} \mathcal{B}_{\phi _{\lambda
			}}-\mathcal{B}_{\phi _0} \mathcal{L}_{\lambda }
			\mathcal{E}_{b_{\phi }}
			\mathcal{B}_{r_{b_r}}-\mathcal{B}_{r_0} \mathcal{L}_{\lambda }
			\mathcal{E}_{b_r} \mathcal{B}_{\phi {b_\phi
			}}\\&+\mathcal{B}_{r_0} \mathcal{L}_{\lambda } \mathcal{E}_{b_{\phi
			}} \mathcal{B}_{\phi_{b_r}}+\mathcal{B}_{r_0}
			\mathcal{E}_{\lambda } \mathcal{L}_{b_r} \mathcal{B}_{\phi
				{b_\phi }}+\mathcal{B}_{r_{\lambda }} \left(\mathcal{L}_0
			\mathcal{E}_{b_r} \mathcal{B}_{\phi {b_\phi
			}}-\mathcal{L}_0 \mathcal{E}_{b_{\phi }} \mathcal{B}_{\phi
				{b_r}}-\mathcal{E}_0 \mathcal{L}_{b_r} \mathcal{B}_{\phi
				{b_\phi }}+\mathcal{E}_0 \mathcal{L}_{b_{\phi }}
			\mathcal{B}_{\phi_{b_r}}+\mathcal{B}_{\phi _0}
			\mathcal{L}_{b_r} \mathcal{E}_{b_{\phi }}-\mathcal{B}_{\phi _0}
			\mathcal{E}_{b_r} \mathcal{L}_{b_{\phi }}\right)\\&-\mathcal{B}_{r_0}
			\mathcal{L}_{b_r} \mathcal{E}_{b_{\phi }} \mathcal{B}_{\phi
				_{\lambda }},
		\end{aligned}
	\end{equation*}
	\begin{equation*}
		\begin{aligned}
			&\Theta_{v0}=\mathcal{B}_{r_{b_\phi}} \left(\mathcal{L}_{\lambda }
			\mathcal{E}_{b_r} \mathcal{B}_{\phi _v}-\mathcal{E}_{\lambda }
			\mathcal{L}_{b_r} \mathcal{B}_{\phi _v}+\mathcal{B}_{\phi
				_{\lambda }} \left(\mathcal{E}_v \mathcal{L}_{b_r}-\mathcal{L}_v
			\mathcal{E}_{b_r}\right)+\mathcal{L}_v \mathcal{E}_{\lambda }
			\mathcal{B}_{\phi_{b_r}}-\mathcal{E}_v \mathcal{L}_{\lambda
			} \mathcal{B}_{\phi_{b_r}}\right)+\mathcal{L}_v
			\mathcal{E}_{\lambda } \left(-\mathcal{B}_{r_{b_r}}\right)
			\mathcal{B}_{\phi_{b_\phi}}+\mathcal{E}_v
			\mathcal{L}_{\lambda } \mathcal{B}_{r_{b_r}}
			\mathcal{B}_{\phi_{b_\phi}}\\&+\mathcal{L}_{b_{\phi }}
			\left(\mathcal{E}_{b_r} \mathcal{B}_{r_v} \mathcal{B}_{\phi
				_{\lambda }}+\mathcal{E}_v \left(-\mathcal{B}_{r_{b_r}}\right)
			\mathcal{B}_{\phi _{\lambda }}+\mathcal{E}_{\lambda }
			\mathcal{B}_{r_{b_r}} \mathcal{B}_{\phi
				_v}-\mathcal{E}_{\lambda } \mathcal{B}_{r_v}
			\mathcal{B}_{\phi_{b_r}}\right)+\mathcal{L}_v
			\mathcal{E}_{b_{\phi }} \mathcal{B}_{r_{b_r}}
			\mathcal{B}_{\phi _{\lambda }}-\mathcal{L}_{\lambda }
			\mathcal{E}_{b_{\phi }} \mathcal{B}_{r_{b_r}}
			\mathcal{B}_{\phi _v}-\mathcal{L}_{\lambda } \mathcal{E}_{b_r}
			\mathcal{B}_{r_v} \mathcal{B}_{\phi {b_\phi
			}}\\&+\mathcal{L}_{\lambda } \mathcal{E}_{b_{\phi }}
			\mathcal{B}_{r_v} \mathcal{B}_{\phi
				{b_r}}+\mathcal{E}_{\lambda } \mathcal{L}_{b_r} \mathcal{B}_{r_v}
			\mathcal{B}_{\phi_{b_\phi}}+\mathcal{B}_{r_{\lambda }}
			\left(\mathcal{L}_v \mathcal{E}_{b_r} \mathcal{B}_{\phi
				{b_\phi }}-\mathcal{L}_v \mathcal{E}_{b_{\phi }}
			\mathcal{B}_{\phi_{b_r}}-\mathcal{E}_v \mathcal{L}_{b_r}
			\mathcal{B}_{\phi_{b_\phi}}+\mathcal{E}_v
			\mathcal{L}_{b_{\phi }} \mathcal{B}_{\phi
				{b_r}}+\mathcal{L}_{b_r} \mathcal{E}_{b_{\phi }} \mathcal{B}_{\phi
				_v}-\mathcal{E}_{b_r} \mathcal{L}_{b_{\phi }} \mathcal{B}_{\phi
				_v}\right)\\&-\mathcal{L}_{b_r} \mathcal{E}_{b_{\phi }}
			\mathcal{B}_{r_v} \mathcal{B}_{\phi _{\lambda }},
		\end{aligned}
	\end{equation*}
	\begin{equation*}
		\begin{aligned}
			&\lambda_{c0}=\mathcal{L}_0 \mathcal{E}_{b_r} \mathcal{B}_{r_{b_\phi}}
			\mathcal{B}_{\phi _{\Theta }}-\mathcal{L}_0 \mathcal{E}_{b_{\phi
			}} \mathcal{B}_{r_{b_r}} \mathcal{B}_{\phi _{\Theta
			}}-\mathcal{B}_{\phi _0} \mathcal{L}_{\Theta } \mathcal{E}_{b_r}
			\mathcal{B}_{r_{b_\phi}}+\mathcal{B}_{\phi _0}
			\mathcal{L}_{\Theta } \mathcal{E}_{b_{\phi }}
			\mathcal{B}_{r_{b_r}}-\mathcal{E}_0 \mathcal{L}_{b_r}
			\mathcal{B}_{r_{b_\phi}} \mathcal{B}_{\phi _{\Theta
			}}+\mathcal{B}_{\phi _0} \mathcal{E}_{\Theta } \mathcal{L}_{b_r}
			\mathcal{B}_{r_{b_\phi}}+\mathcal{E}_0 \mathcal{L}_{b_{\phi
			}} \mathcal{B}_{r_{b_r}} \mathcal{B}_{\phi _{\Theta
			}}\\&-\mathcal{B}_{\phi _0} \mathcal{E}_{\Theta }
			\mathcal{L}_{b_{\phi }} \mathcal{B}_{r_{b_r}}-\mathcal{L}_0
			\mathcal{E}_{b_r} \mathcal{B}_{r_{\Theta }}
			\mathcal{B}_{\phi_{b_\phi}}+\mathcal{L}_0
			\mathcal{E}_{b_{\phi }} \mathcal{B}_{r_{\Theta }}
			\mathcal{B}_{\phi_{b_r}}+\mathcal{B}_{r_0}
			\mathcal{L}_{\Theta } \mathcal{E}_{b_r} \mathcal{B}_{\phi
				{b_\phi }}-\mathcal{B}_{r_0} \mathcal{L}_{\Theta }
			\mathcal{E}_{b_{\phi }} \mathcal{B}_{\phi
				{b_r}}+\mathcal{E}_0 \mathcal{L}_{b_r} \mathcal{B}_{r_{\Theta }}
			\mathcal{B}_{\phi_{b_\phi}}-\mathcal{B}_{r_0}
			\mathcal{E}_{\Theta } \mathcal{L}_{b_r} \mathcal{B}_{\phi
				{b_\phi }}\\&-\mathcal{E}_0 \mathcal{L}_{b_{\phi }}
			\mathcal{B}_{r_{\Theta }} \mathcal{B}_{\phi
				{b_r}}+\mathcal{B}_{r_0} \mathcal{E}_{\Theta }
			\mathcal{L}_{b_{\phi }} \mathcal{B}_{\phi
				{b_r}}-\mathcal{B}_{\phi _0} \mathcal{L}_{b_r}
			\mathcal{E}_{b_{\phi }} \mathcal{B}_{r_{\Theta
			}}+\mathcal{B}_{r_0} \mathcal{L}_{b_r} \mathcal{E}_{b_{\phi }}
			\mathcal{B}_{\phi _{\Theta }}+\mathcal{B}_{\phi _0}
			\mathcal{E}_{b_r} \mathcal{L}_{b_{\phi }} \mathcal{B}_{r_{\Theta
			}}-\mathcal{B}_{r_0} \mathcal{E}_{b_r} \mathcal{L}_{b_{\phi }}
			\mathcal{B}_{\phi _{\Theta }}+\mathcal{L}_0 \mathcal{E}_{\Theta }
			\mathcal{B}_{\phi_{b_r}}
			\left(-\mathcal{B}_{r_{b_\phi}}\right)\\&+\mathcal{L}_0
			\mathcal{E}_{\Theta } \mathcal{B}_{r_{b_r}}
			\mathcal{B}_{\phi_{b_\phi}}+\mathcal{E}_0
			\mathcal{L}_{\Theta } \mathcal{B}_{\phi_{b_r}}
			\mathcal{B}_{r_{b_\phi}}-\mathcal{E}_0 \mathcal{L}_{\Theta
			} \mathcal{B}_{r_{b_r}} \mathcal{B}_{\phi_{b_\phi}},
		\end{aligned}
	\end{equation*}
	\begin{equation*}
		\begin{aligned}
			&\lambda_{v0}=\mathcal{L}_v \mathcal{E}_{b_r} \mathcal{B}_{r_{b_\phi}}
			\mathcal{B}_{\phi _{\Theta }}-\mathcal{L}_v \mathcal{E}_{b_{\phi
			}} \mathcal{B}_{r_{b_r}} \mathcal{B}_{\phi _{\Theta
			}}-\mathcal{L}_{\Theta } \mathcal{E}_{b_r}
			\mathcal{B}_{r_{b_\phi}} \mathcal{B}_{\phi
				_v}+\mathcal{L}_{\Theta } \mathcal{E}_{b_{\phi }}
			\mathcal{B}_{r_{b_r}} \mathcal{B}_{\phi _v}-\mathcal{E}_v
			\mathcal{L}_{b_r} \mathcal{B}_{r_{b_\phi}}
			\mathcal{B}_{\phi _{\Theta }}+\mathcal{E}_{\Theta }
			\mathcal{L}_{b_r} \mathcal{B}_{r_{b_\phi}}
			\mathcal{B}_{\phi _v}+\mathcal{E}_v \mathcal{L}_{b_{\phi }}
			\mathcal{B}_{r_{b_r}} \mathcal{B}_{\phi _{\Theta
			}}\\&-\mathcal{E}_{\Theta } \mathcal{L}_{b_{\phi }}
			\mathcal{B}_{r_{b_r}} \mathcal{B}_{\phi _v}-\mathcal{L}_v
			\mathcal{E}_{b_r} \mathcal{B}_{r_{\Theta }}
			\mathcal{B}_{\phi_{b_\phi}}+\mathcal{L}_v
			\mathcal{E}_{b_{\phi }} \mathcal{B}_{r_{\Theta }}
			\mathcal{B}_{\phi_{b_r}}+\mathcal{L}_{\Theta }
			\mathcal{E}_{b_r} \mathcal{B}_{r_v} \mathcal{B}_{\phi
				{b_\phi }}-\mathcal{L}_{\Theta } \mathcal{E}_{b_{\phi }}
			\mathcal{B}_{r_v} \mathcal{B}_{\phi_{b_r}}+\mathcal{E}_v
			\mathcal{L}_{b_r} \mathcal{B}_{r_{\Theta }}
			\mathcal{B}_{\phi_{b_\phi}}-\mathcal{E}_{\Theta }
			\mathcal{L}_{b_r} \mathcal{B}_{r_v} \mathcal{B}_{\phi
				{b_\phi }}\\&-\mathcal{E}_v \mathcal{L}_{b_{\phi }}
			\mathcal{B}_{r_{\Theta }} \mathcal{B}_{\phi
				{b_r}}+\mathcal{E}_{\Theta } \mathcal{L}_{b_{\phi }}
			\mathcal{B}_{r_v} \mathcal{B}_{\phi
				{b_r}}-\mathcal{L}_{b_r} \mathcal{E}_{b_{\phi }}
			\mathcal{B}_{r_{\Theta }} \mathcal{B}_{\phi _v}+\mathcal{L}_{b_r}
			\mathcal{E}_{b_{\phi }} \mathcal{B}_{r_v} \mathcal{B}_{\phi
				_{\Theta }}+\mathcal{E}_{b_r} \mathcal{L}_{b_{\phi }}
			\mathcal{B}_{r_{\Theta }} \mathcal{B}_{\phi _v}-\mathcal{E}_{b_r}
			\mathcal{L}_{b_{\phi }} \mathcal{B}_{r_v} \mathcal{B}_{\phi
				_{\Theta }}+\mathcal{L}_v \mathcal{E}_{\Theta }
			\mathcal{B}_{\phi_{b_r}}
			\left(-\mathcal{B}_{r_{b_\phi}}\right)\\&+\mathcal{L}_v
			\mathcal{E}_{\Theta } \mathcal{B}_{r_{b_r}}
			\mathcal{B}_{\phi_{b_\phi}}+\mathcal{E}_v
			\mathcal{L}_{\Theta } \mathcal{B}_{\phi_{b_r}}
			\mathcal{B}_{r_{b_\phi}}-\mathcal{E}_v \mathcal{L}_{\Theta
			} \mathcal{B}_{r_{b_r}} \mathcal{B}_{\phi_{b_\phi}},
		\end{aligned}
	\end{equation*}
	\begin{equation*}
		\begin{aligned}
			&b^r_{c0}=-\mathcal{L}_0 \mathcal{E}_{b_{\phi }} \mathcal{B}_{r_{\lambda }}
			\mathcal{B}_{\phi _{\Theta }}+\mathcal{L}_0 \mathcal{E}_{b_{\phi
			}} \mathcal{B}_{r_{\Theta }} \mathcal{B}_{\phi _{\lambda
			}}+\mathcal{B}_{\phi _0} \mathcal{L}_{\Theta }
			\mathcal{E}_{b_{\phi }} \mathcal{B}_{r_{\lambda
			}}-\mathcal{B}_{r_0} \mathcal{L}_{\Theta } \mathcal{E}_{b_{\phi }}
			\mathcal{B}_{\phi _{\lambda }}-\mathcal{B}_{\phi _0}
			\mathcal{L}_{\lambda } \mathcal{E}_{b_{\phi }}
			\mathcal{B}_{r_{\Theta }}+\mathcal{B}_{r_0} \mathcal{L}_{\lambda }
			\mathcal{E}_{b_{\phi }} \mathcal{B}_{\phi _{\Theta
			}}+\mathcal{E}_0 \mathcal{L}_{b_{\phi }} \mathcal{B}_{r_{\lambda
			}} \mathcal{B}_{\phi _{\Theta }}\\&-\mathcal{E}_0
			\mathcal{L}_{b_{\phi }} \mathcal{B}_{r_{\Theta }}
			\mathcal{B}_{\phi _{\lambda }}-\mathcal{B}_{\phi _0}
			\mathcal{E}_{\Theta } \mathcal{L}_{b_{\phi }}
			\mathcal{B}_{r_{\lambda }}+\mathcal{B}_{r_0} \mathcal{E}_{\Theta }
			\mathcal{L}_{b_{\phi }} \mathcal{B}_{\phi _{\lambda
			}}+\mathcal{B}_{\phi _0} \mathcal{E}_{\lambda }
			\mathcal{L}_{b_{\phi }} \mathcal{B}_{r_{\Theta
			}}-\mathcal{B}_{r_0} \mathcal{E}_{\lambda } \mathcal{L}_{b_{\phi
			}} \mathcal{B}_{\phi _{\Theta }}+\mathcal{L}_0 \mathcal{E}_{\Theta
			} \mathcal{B}_{r_{\lambda }} \mathcal{B}_{\phi {b_\phi
			}}-\mathcal{L}_0 \mathcal{E}_{\lambda } \mathcal{B}_{r_{\Theta }}
			\mathcal{B}_{\phi_{b_\phi}}\\&-\mathcal{E}_0
			\mathcal{L}_{\Theta } \mathcal{B}_{r_{\lambda }}
			\mathcal{B}_{\phi_{b_\phi}}+\mathcal{B}_{r_0}
			\mathcal{E}_{\lambda } \mathcal{L}_{\Theta }
			\mathcal{B}_{\phi_{b_\phi}}+\mathcal{E}_0
			\mathcal{L}_{\lambda } \mathcal{B}_{r_{\Theta }}
			\mathcal{B}_{\phi_{b_\phi}}-\mathcal{B}_{r_0}
			\mathcal{E}_{\Theta } \mathcal{L}_{\lambda }
			\mathcal{B}_{\phi_{b_\phi}}+\mathcal{L}_0
			\mathcal{E}_{\Theta } \left(-\mathcal{B}_{r_{b_\phi
			}}\right) \mathcal{B}_{\phi _{\lambda }}+\mathcal{L}_0
			\mathcal{E}_{\lambda } \mathcal{B}_{r_{b_\phi}}
			\mathcal{B}_{\phi _{\Theta }}+\mathcal{E}_0 \mathcal{L}_{\Theta }
			\mathcal{B}_{r_{b_\phi}} \mathcal{B}_{\phi _{\lambda
			}}\\&-\mathcal{B}_{\phi _0} \mathcal{E}_{\lambda }
			\mathcal{L}_{\Theta } \mathcal{B}_{r_{b_\phi
			}}-\mathcal{E}_0 \mathcal{L}_{\lambda }
			\mathcal{B}_{r_{b_\phi}} \mathcal{B}_{\phi _{\Theta
			}}+\mathcal{B}_{\phi _0} \mathcal{E}_{\Theta }
			\mathcal{L}_{\lambda } \mathcal{B}_{r_{b_\phi}},
		\end{aligned}
	\end{equation*}
	\begin{equation*}
		\begin{aligned}
			&b^r_{v0}=-\mathcal{L}_v \mathcal{E}_{b_{\phi }} \mathcal{B}_{r_{\lambda }}
			\mathcal{B}_{\phi _{\Theta }}+\mathcal{L}_v \mathcal{E}_{b_{\phi
			}} \mathcal{B}_{r_{\Theta }} \mathcal{B}_{\phi _{\lambda
			}}+\mathcal{L}_{\Theta } \mathcal{E}_{b_{\phi }}
			\mathcal{B}_{r_{\lambda }} \mathcal{B}_{\phi
				_v}-\mathcal{L}_{\Theta } \mathcal{E}_{b_{\phi }}
			\mathcal{B}_{r_v} \mathcal{B}_{\phi _{\lambda
			}}-\mathcal{L}_{\lambda } \mathcal{E}_{b_{\phi }}
			\mathcal{B}_{r_{\Theta }} \mathcal{B}_{\phi
				_v}+\mathcal{L}_{\lambda } \mathcal{E}_{b_{\phi }}
			\mathcal{B}_{r_v} \mathcal{B}_{\phi _{\Theta }}+\mathcal{E}_v
			\mathcal{L}_{b_{\phi }} \mathcal{B}_{r_{\lambda }}
			\mathcal{B}_{\phi _{\Theta }}\\&-\mathcal{E}_v \mathcal{L}_{b_{\phi
			}} \mathcal{B}_{r_{\Theta }} \mathcal{B}_{\phi _{\lambda
			}}-\mathcal{E}_{\Theta } \mathcal{L}_{b_{\phi }}
			\mathcal{B}_{r_{\lambda }} \mathcal{B}_{\phi
				_v}+\mathcal{E}_{\Theta } \mathcal{L}_{b_{\phi }}
			\mathcal{B}_{r_v} \mathcal{B}_{\phi _{\lambda
			}}+\mathcal{E}_{\lambda } \mathcal{L}_{b_{\phi }}
			\mathcal{B}_{r_{\Theta }} \mathcal{B}_{\phi
				_v}-\mathcal{E}_{\lambda } \mathcal{L}_{b_{\phi }}
			\mathcal{B}_{r_v} \mathcal{B}_{\phi _{\Theta }}+\mathcal{L}_v
			\mathcal{E}_{\Theta } \mathcal{B}_{r_{\lambda }}
			\mathcal{B}_{\phi_{b_\phi}}-\mathcal{L}_v
			\mathcal{E}_{\lambda } \mathcal{B}_{r_{\Theta }}
			\mathcal{B}_{\phi_{b_\phi}}\\&-\mathcal{E}_v
			\mathcal{L}_{\Theta } \mathcal{B}_{r_{\lambda }}
			\mathcal{B}_{\phi_{b_\phi}}+\mathcal{E}_{\lambda }
			\mathcal{L}_{\Theta } \mathcal{B}_{r_v} \mathcal{B}_{\phi
				{b_\phi }}+\mathcal{E}_v \mathcal{L}_{\lambda }
			\mathcal{B}_{r_{\Theta }} \mathcal{B}_{\phi {b_\phi
			}}-\mathcal{E}_{\Theta } \mathcal{L}_{\lambda } \mathcal{B}_{r_v}
			\mathcal{B}_{\phi_{b_\phi}}+\mathcal{L}_v
			\mathcal{E}_{\Theta } \left(-\mathcal{B}_{r {b_\phi
			}}\right) \mathcal{B}_{\phi _{\lambda }}+\mathcal{L}_v
			\mathcal{E}_{\lambda } \mathcal{B}_{r_{b_\phi}}
			\mathcal{B}_{\phi _{\Theta }}+\mathcal{E}_v \mathcal{L}_{\Theta }
			\mathcal{B}_{r_{b_\phi}} \mathcal{B}_{\phi _{\lambda
			}}\\&-\mathcal{E}_{\lambda } \mathcal{L}_{\Theta }
			\mathcal{B}_{r_{b_\phi}} \mathcal{B}_{\phi
				_v}-\mathcal{E}_v \mathcal{L}_{\lambda }
			\mathcal{B}_{r_{b_\phi}} \mathcal{B}_{\phi _{\Theta
			}}+\mathcal{E}_{\Theta } \mathcal{L}_{\lambda }
			\mathcal{B}_{r_{b_\phi}} \mathcal{B}_{\phi _v},
		\end{aligned}
	\end{equation*}
	\begin{equation*}
		\begin{aligned}
			&b^{\phi}_{c0}=\mathcal{L}_0 \mathcal{E}_{b_r} \mathcal{B}_{r_{\lambda }}
			\mathcal{B}_{\phi _{\Theta }}-\mathcal{L}_0 \mathcal{E}_{b_r}
			\mathcal{B}_{r_{\Theta }} \mathcal{B}_{\phi _{\lambda
			}}-\mathcal{B}_{\phi _0} \mathcal{L}_{\Theta } \mathcal{E}_{b_r}
			\mathcal{B}_{r_{\lambda }}+\mathcal{B}_{r_0} \mathcal{L}_{\Theta }
			\mathcal{E}_{b_r} \mathcal{B}_{\phi _{\lambda }}+\mathcal{B}_{\phi
				_0} \mathcal{L}_{\lambda } \mathcal{E}_{b_r}
			\mathcal{B}_{r_{\Theta }}-\mathcal{B}_{r_0} \mathcal{L}_{\lambda }
			\mathcal{E}_{b_r} \mathcal{B}_{\phi _{\Theta }}-\mathcal{E}_0
			\mathcal{L}_{b_r} \mathcal{B}_{r_{\lambda }} \mathcal{B}_{\phi
				_{\Theta }}\\&+\mathcal{E}_0 \mathcal{L}_{b_r} \mathcal{B}_{r_{\Theta
			}} \mathcal{B}_{\phi _{\lambda }}+\mathcal{B}_{\phi _0}
			\mathcal{E}_{\Theta } \mathcal{L}_{b_r} \mathcal{B}_{r_{\lambda
			}}-\mathcal{B}_{r_0} \mathcal{E}_{\Theta } \mathcal{L}_{b_r}
			\mathcal{B}_{\phi _{\lambda }}-\mathcal{B}_{\phi _0}
			\mathcal{E}_{\lambda } \mathcal{L}_{b_r} \mathcal{B}_{r_{\Theta
			}}+\mathcal{B}_{r_0} \mathcal{E}_{\lambda } \mathcal{L}_{b_r}
			\mathcal{B}_{\phi _{\Theta }}+\mathcal{L}_0 \mathcal{E}_{\Theta }
			\mathcal{B}_{r_{b_r}} \mathcal{B}_{\phi _{\lambda
			}}-\mathcal{L}_0 \mathcal{E}_{\lambda } \mathcal{B}_{r_{b_r}}
			\mathcal{B}_{\phi _{\Theta }}\\&-\mathcal{E}_0 \mathcal{L}_{\Theta }
			\mathcal{B}_{r_{b_r}} \mathcal{B}_{\phi _{\lambda
			}}+\mathcal{B}_{\phi _0} \mathcal{E}_{\lambda }
			\mathcal{L}_{\Theta } \mathcal{B}_{r_{b_r}}+\mathcal{E}_0
			\mathcal{L}_{\lambda } \mathcal{B}_{r_{b_r}} \mathcal{B}_{\phi
				_{\Theta }}-\mathcal{B}_{\phi _0} \mathcal{E}_{\Theta }
			\mathcal{L}_{\lambda } \mathcal{B}_{r_{b_r}}-\mathcal{L}_0
			\mathcal{E}_{\Theta } \mathcal{B}_{r_{\lambda }}
			\mathcal{B}_{\phi_{b_r}}+\mathcal{L}_0 \mathcal{E}_{\lambda
			} \mathcal{B}_{r_{\Theta }} \mathcal{B}_{\phi
				{b_r}}\\&+\mathcal{E}_0 \mathcal{L}_{\Theta } \mathcal{B}_{r_{\lambda
			}} \mathcal{B}_{\phi_{b_r}}-\mathcal{B}_{r_0}
			\mathcal{E}_{\lambda } \mathcal{L}_{\Theta }
			\mathcal{B}_{\phi_{b_r}}-\mathcal{E}_0 \mathcal{L}_{\lambda
			} \mathcal{B}_{r_{\Theta }} \mathcal{B}_{\phi
				{b_r}}+\mathcal{B}_{r_0} \mathcal{E}_{\Theta }
			\mathcal{L}_{\lambda } \mathcal{B}_{\phi_{b_r}},
		\end{aligned}
	\end{equation*}
	\begin{equation*}
		\begin{aligned}
			&b^{\phi}_{v0}=\mathcal{L}_v \mathcal{E}_{b_r} \mathcal{B}_{r_{\lambda }}
			\mathcal{B}_{\phi _{\Theta }}-\mathcal{L}_v \mathcal{E}_{b_r}
			\mathcal{B}_{r_{\Theta }} \mathcal{B}_{\phi _{\lambda
			}}-\mathcal{L}_{\Theta } \mathcal{E}_{b_r} \mathcal{B}_{r_{\lambda
			}} \mathcal{B}_{\phi _v}+\mathcal{L}_{\Theta } \mathcal{E}_{b_r}
			\mathcal{B}_{r_v} \mathcal{B}_{\phi _{\lambda
			}}+\mathcal{L}_{\lambda } \mathcal{E}_{b_r} \mathcal{B}_{r_{\Theta
			}} \mathcal{B}_{\phi _v}-\mathcal{L}_{\lambda } \mathcal{E}_{b_r}
			\mathcal{B}_{r_v} \mathcal{B}_{\phi _{\Theta }}-\mathcal{E}_v
			\mathcal{L}_{b_r} \mathcal{B}_{r_{\lambda }} \mathcal{B}_{\phi
				_{\Theta }}\\&+\mathcal{E}_v \mathcal{L}_{b_r} \mathcal{B}_{r_{\Theta
			}} \mathcal{B}_{\phi _{\lambda }}+\mathcal{E}_{\Theta }
			\mathcal{L}_{b_r} \mathcal{B}_{r_{\lambda }} \mathcal{B}_{\phi
				_v}-\mathcal{E}_{\Theta } \mathcal{L}_{b_r} \mathcal{B}_{r_v}
			\mathcal{B}_{\phi _{\lambda }}-\mathcal{E}_{\lambda }
			\mathcal{L}_{b_r} \mathcal{B}_{r_{\Theta }} \mathcal{B}_{\phi
				_v}+\mathcal{E}_{\lambda } \mathcal{L}_{b_r} \mathcal{B}_{r_v}
			\mathcal{B}_{\phi _{\Theta }}+\mathcal{L}_v \mathcal{E}_{\Theta }
			\mathcal{B}_{r_{b_r}} \mathcal{B}_{\phi _{\lambda
			}}-\mathcal{L}_v \mathcal{E}_{\lambda } \mathcal{B}_{r_{b_r}}
			\mathcal{B}_{\phi _{\Theta }}\\&-\mathcal{E}_v \mathcal{L}_{\Theta }
			\mathcal{B}_{r_{b_r}} \mathcal{B}_{\phi _{\lambda
			}}+\mathcal{E}_{\lambda } \mathcal{L}_{\Theta }
			\mathcal{B}_{r_{b_r}} \mathcal{B}_{\phi _v}+\mathcal{E}_v
			\mathcal{L}_{\lambda } \mathcal{B}_{r_{b_r}} \mathcal{B}_{\phi
				_{\Theta }}-\mathcal{E}_{\Theta } \mathcal{L}_{\lambda }
			\mathcal{B}_{r_{b_r}} \mathcal{B}_{\phi _v}-\mathcal{L}_v
			\mathcal{E}_{\Theta } \mathcal{B}_{r_{\lambda }}
			\mathcal{B}_{\phi_{b_r}}+\mathcal{L}_v \mathcal{E}_{\lambda
			} \mathcal{B}_{r_{\Theta }} \mathcal{B}_{\phi_{b_r}}+\mathcal{E}_v \mathcal{L}_{\Theta } \mathcal{B}_{r_{\lambda
			}} \mathcal{B}_{\phi_{b_r}}\\&-\mathcal{E}_{\lambda }
			\mathcal{L}_{\Theta } \mathcal{B}_{r_v} \mathcal{B}_{\phi
				{b_r}}-\mathcal{E}_v \mathcal{L}_{\lambda } \mathcal{B}_{r_{\Theta
			}} \mathcal{B}_{\phi_{b_r}}+\mathcal{E}_{\Theta }
			\mathcal{L}_{\lambda } \mathcal{B}_{r_v} \mathcal{B}_{\phi
				{b_r}}.
		\end{aligned}
	\end{equation*}

\end{document}